\documentclass[nofootinbib,preprintnumbers,prd,superscriptaddress,twocolumn]{revtex4}

\usepackage{amsmath}
\usepackage{amsfonts}
\usepackage{amssymb}
\usepackage{graphicx}
\usepackage[titletoc]{appendix}
\usepackage{color}
\usepackage{hyperref}
\usepackage{cleveref}
\usepackage[rightcaption]{sidecap}
\usepackage{subfigure}
\usepackage{comment}

\usepackage{dcolumn}

\usepackage{array}
\usepackage{ctable}
\usepackage{multirow}
\usepackage{siunitx}
\usepackage{longtable}
\usepackage{tabularx}
\usepackage{booktabs}

\usepackage{amsthm}
\usepackage{mathrsfs}

\usepackage{epsfig}
\usepackage{amscd}

\usepackage{bm}
\usepackage{natbib}
\usepackage{url}
\usepackage{xspace}
\usepackage{kantlipsum}

\graphicspath{{Graphics/}}

\def\be{\begin{equation}}
\def\ee{\end{equation}}
\def\bea{\begin{eqnarray}}
\def\eea{\end{eqnarray}}

\def\nat{Nature}
\def\prl{Phys. Rev. Lett.}

\def\prd{Phys. Rev. D}

\def\mnras{MNRAS}
\def\apj{ApJ}
\def\apjl{ApJ Lett.}

\def\aap{A\&A}

\def\pasj{Publ. Astr. Soc. Japan }
\def\pasp{Publ. Astr. Soc. Pacific}

\def\nphysa{Nucl. Phys.}

\def\jcap{JCAP}

\def\nar{New Astron. Rev.}

\definecolor{vividviolet}{rgb}{0.62, 0.0, 1.0}
\definecolor{amaranth}{rgb}{0.9, 0.17, 0.31}
\definecolor{palatinateblue}{rgb}{0.15, 0.23, 0.89}
\definecolor{brightpink}{rgb}{1.0, 0.0, 0.5}
\definecolor{cornflowerblue}{rgb}{0.39, 0.58, 0.93}
\definecolor{deepcarminepink}{rgb}{0.94, 0.19, 0.22}
\definecolor{radicalred}{rgb}{1.0, 0.21, 0.37}
\hypersetup{ linktoc=all,
    colorlinks, linkcolor={palatinateblue},
    citecolor={brightpink}, urlcolor={amaranth}
}

\begin{document}

\title{Quasi-periodic oscillations for spherically symmetric regular black holes}

\author{Kuantay~Boshkayev}
\email{kuantay@mail.ru}
\affiliation{NNLOT, Al-Farabi Kazakh National University, Al-Farabi av. 71, 050040 Almaty, Kazakhstan.}
\affiliation{International University of Information Technology, Manas st. 34/1, 050040 Almaty, Kazakhstan.}

\author{Anuar~Idrissov}
\email{anuar.idrissov@gmail.com}
\affiliation{Department of Physics, Nazarbayev University, Kabanbay Batyr 53, 010000 Astana, Kazakhstan.}

\author{Orlando~Luongo}
\email{orlando.luongo@unicam.it}
\affiliation{NNLOT, Al-Farabi Kazakh National University, Al-Farabi av. 71, 050040 Almaty, Kazakhstan.}
\affiliation{Universit\`a di Camerino, Via Madonna delle Carceri 9, 62032 Camerino, Italy.}
\affiliation{SUNY Polytechnic Institute, 13502 Utica, New York, USA.}
\affiliation{Istituto Nazionale di Fisica Nucleare, Sezione di Perugia, 06123, Perugia,  Italy.}

\author{Marco Muccino}
\email{muccino@lnf.infn.it}
\affiliation{NNLOT, Al-Farabi Kazakh National University, Al-Farabi av. 71, 050040 Almaty, Kazakhstan.}

\begin{abstract}
We consider the 
recent data sets of quasi-periodic oscillations from eight different low mass X-ray binaries.
We here interpret their physical features in the context of given regular black hole solutions and verify their applicability to neutron star configurations.
We evaluate the numerical constraints over the free parameters of Bardeen, Hayward and Dymnikova regular solutions by performing a set of Markov chain Monte Carlo analyses, based on the Metropolis algorithm.
For each source, we evaluate the best-fit parameters, among which mass and magnetic charge, and compare and contrast them with the current literature.
We also infer the corresponding innermost stable circular orbit radii and the radial extents of the accretion disks.
Focusing on how to identify discrepancies among theoretical models and observations, our results show that, in most of the cases, regular black holes, in particular the Bardeen and Hayward spacetimes are slightly more suitable to describe neutron stars than Schwarzschild geometry, whereas the Dymnikova metric is ruled out.
\end{abstract}

\maketitle

\section{Introduction}

Recently, a novel interest in \emph{black hole} (BH) physics arose due to the detection of \emph{gravitational waves} \cite{2016PhRvL.116f1102A} first and then to the impressive discovery of \emph{BH shadows} \cite{2019ApJ...875L...1E}, culminating to new confirmations of Einstein's general relativity\footnote{For clarity, from the above breakthroughs we only argued tighter bounds on extended and/or modified theories of gravity \cite{Astashenok:2014nua,Astashenok:2013vza,Astashenok:2017dpo,Capozziello:2019cav,Astashenok:2020qds}. However, no definitive conclusions are reached in regimes where gravity breaks down, i.e., quantum gravity epoch, strong field regimes, and so on.} (GR). These findings determine a current understanding of BHs and may shed light on how, and whether, Einstein's gravity fails to be predictive. Consequently, all these advancements have led to the beginning of a new era based on \emph{BH precision astronomy} \cite{2021NatRP...3..732V}.

In this fascinating scenario, low-mass X-ray binaries and microquasars may play a complementary role. These sources exhibit narrow peaks of excess energy in their X-ray fluxes, named \emph{quasi-periodic oscillations} (QPOs), that are associated with matter accretion into compact objects, providing information about the system and its overall properties. Analogously, they can be adopted to detect possible extensions of Einstein's gravity, becoming essential to furnish new data in the framework of BH astronomy. Such oscillations may conventionally be classified into two main typologies: \emph{high-frequencies}, say $0.1-1$ kHz, and \emph{low frequencies}, say below $0.1$ kHz.

For instance, some of them suggest that the high frequency are related to the Keplerian frequency close to  the innermost stable circular orbit (ISCO) of test particles  in accretion disks \cite{2006csxs.book...39V, Lamb2008}, whereas low frequencies are normally associated with periastron precession frequency \cite{1999NuPhS..69..135S,2014GrCo...20..233B,2015ARep...59..441B, 2016ApJ...833..273T, 2018mgm..conf.3433B}.

In the literature, there is a \emph{plethora} of models describing QPOs data \cite{1998ApJ...508..791M, 1998ApJ...499..315T, 2001ApJ...554.1210L, 2001ApJ...553..955F, 2001A&A...374L..19A, 2002ApJ...577L..23T, 2004PASJ...56..553R, 2005PASJ...57L..17K, 2012PASJ...64..139K, 2022JCAP...05..020B}. The most-accredited paradigm is the \emph{relativistic precession model}, where QPOs emerge from the motion of inhomogeneities in accretion mechanism \cite{1998ApJ...492L..59S, 1999PhRvL..82...17S,1999ApJ...524L..63S, 2022MNRAS.517.1469M,2022MNRAS.517.1389R}. On the other hand, in the framework of alternative theories of gravity, it is also possible to describe compact objects producing QPOs, however involving GR approaches \cite{2022arXiv221210186B}. Finally, BH binaries may also exhibit QPOs, again classified into low and high frequencies \cite{2022MNRAS.510..807S,2022MNRAS.515.2099B, 2022PASJ...74.1220S,2022MNRAS.514.2891Z}. Consequently, a definitive physical mechanism behind QPOs is not fully-understood yet, since their description can be intimately related to astrophysical BHs, whose theoretical features are currently under investigation \cite{2019NewAR..8501524I,motta2016quasi}.

In this puzzle, by virtue of Penrose and Hawking singularity theorems \cite{1965PhRvL..14...57P, 1970RSPSA.314..529H}, the presence of matter satisfying reasonable energy conditions inevitably leads to singularities in GR. However, it is widely-believed that such singularities can be healed introducing a complete theory of \emph{quantum gravity}.

Thus, attempting to classically-overcome the above issue, Bardeen introduced the concept of a regular BH (RBH) \cite{bardeen1968proceedings}, where a non-singular center occurs, exhibiting asymptotic flatness, in a static spherical symmetry. Later, Ayón-Beato and García \cite{2000PhLB..493..149A}  demonstrated that it could be considered as a solution of Einstein's field equations coupled with a magnetic monopole source in the context of nonlinear electrodynamics \cite{1998PhRvL..80.5056A}. Subsequently, other RBH models were proposed, e.g. Dymnikova \cite{1992GReGr..24..235D} introduced a different type of RBH coinciding with the Schwarzschild spacetime at infinity and behaving as a de Sitter solution close to the center. Additionally, Hayward proposed a static and spherically symmetric RBH model to fulfill the BH information-loss paradox \cite{2006PhRvL..96c1103H}.

Assuming that RBH solutions may also model neutron stars (NSs), we here investigate whether RBHs can be used in framing out sources of experimental QPO data, where the effects of (topological) charge cannot be excluded \emph{a priori}. To do so, following recent studies, supporting the relativistic precession model, we hereafter associate QPO modes with the the fundamental frequencies provided by a test-particle in the background metric, i.e., azimuthal (or Keplerian) frequency along with radial and vertical epicyclic frequencies. Thus, employing spherical symmetry, we work out the Bardeen, Hayward and Dymnikova spacetimes, and fit these metrics with the QPO data catalogs. In so doing, we check whether it is plausible to use those configurations with the aim of describing the exterior of NSs. Specifically, the first two metrics assume the existence of topological charge; only the third is constructed without this assumption.

Thus, in testing such solutions, we involve eight NS sources in the low mass X-ray binaries, considering their most updated QPO data sets. To test them, we numerically maximize the corresponding log-likelihoods, performing Markov chain Monte Carlo (MCMC) analyses, based on the \emph{Metropolis-Hastings} algorithm, aiming to get best-fit parameters and the direct $1$--$\sigma$ and $2$--$\sigma$ error bars for each RBH solution. Guided by the NS interpretation and by the values of the ISCO of each metric, we show that in most cases RBHs are successful to model QPO frequencies. For the majority of our sources, we therefore end up that the Schwarzschild metric does not remain the unique space-time able to provide physical predictions from such sources. In particular, RBHs  are favored as byproduct of our physical scrutiny. Among our choices, we see that the Bardeen and Hayward solutions appear favored, whereas the Dymnikova metric is \emph{ruled out}. Physical consequences of our findings are also debated, comparing our outcomes with previous results, available in the literature.

The paper is organized as follows. In Sec.~II, we describe QPOs and the associated frequencies in the framework of the relativistic precession model. In Sec.~III we introduce the proposed regular spacetimes, highlighting their physical main features. In Sec.~IV, we perform MCMC analyses and, in Sec.~V, we discuss the physical interpretations and their theoretical implications. In Sec.~VI, we report conclusions and perspectives of the work\footnote{Throughout this paper we use natural units, $G=c= \hbar = 1$, and Lorentzian  signature $(-,+,+,+)$.}.


\section{Quasi-periodic oscillations}

QPOs are  not strictly periodic oscillations, prompting a slightly varying set of frequencies that can be viewed as nearly constant. These frequencies are often observed in astrophysical systems, among which  accretion disks around BHs, NSs and, in a broad sense, from strong gravity compact objects.

QPOs were first discovered in white dwarfs \cite{1982ApJ...257L..71M, 1987A&A...181L..15L, 1989A&A...217..146L,2015A&A...579A..24B} and later in the power spectra of X-ray binary NSs and BHs, leading to further studies of accretion disks \cite{1985Natur.316..225V, 1986ApJ...306..230M, 2005A&A...440L..25K, 2007PASP..119..393Z, 2012MNRAS.426.1701B, 2014PhRvD..89l7302B}. QPOs provide a way to test gravity and to gather information about sources and cosmology \cite{2009JCAP...09..013B}. Tight QPO frequency measurements obtained from accretion disks around compact objects help to determine the most suitable model for astrophysical processes in these systems \cite{2011CQGra..28k4009M, 2020EPJC...80..133K}.

The harmonic oscillation we refer to is related to the fundamental or epicyclic frequencies of test particles moving in circular orbits within accretion disks around compact objects, so that the corresponding procedure to estimate those frequencies  starts through a test particle Lagrangian
\begin{equation}
    \mathbb L = \frac{1}{2} m g_{\mu\nu} \dot{x}^{\mu} \dot{x}^{\nu}\,,
\end{equation}
where $m$ is the test particle mass and $x^{\mu}$ are four-coordinates with $\mu =0,1,2,3$. Here, $\dot{x}^{\mu}=d x^{\mu}/d \tau $ represent their four-velocities.
Since we have static metrics, the killing vectors imply conserved quantities, say the specific energy, $\mathcal E = -g_{tt}\dot{t}$, and the angular momentum, $\mathcal L = g_{\phi \phi} \dot{\phi}$. For massive test particles, the four-velocity normalization condition is
$g_{\mu\nu} \dot{x}^{\mu} \dot{x}^{\nu}=-1$, from which we require to get its radial component.

Thus, as $m \neq 0$, the integrals of motion read
\begin{equation}
\dot{t} = - \frac{\cal E}{g_{tt}}\,, \quad\quad \\
\dot{\phi} = \frac{\cal L}{g_{\phi \phi}}\,,  \quad\quad \\
g_{rr} \dot{r}^2 + g_{\theta\theta} \dot{\theta}^2 = V_{eff}\,,
\end{equation}
where the effective potential, $V_{eff}$, is defined as
\begin{equation}
 V_{eff} = - 1 - \frac{{\cal E}^2 g_{\phi \phi} + {\cal L}^2 g_{tt}}{ g_{tt} g_{\phi \phi} }\,.
\end{equation}
For circular orbits in the equatorial plane one has $\dot{r}=\dot{\theta}=0$ and, so, the equations for orbital parameters of test particles are given by
\begin{subequations}
\begin{align}
\Omega _\phi & = \pm \sqrt{-\frac{\partial_r g_{tt}}{\partial_r g_{\phi \phi}}}\,, \\
\dot{t} &= u^t = \frac{1}{\sqrt{-g_{tt}-g_{\phi \phi} \Omega^2_{\phi}}}\,,  \\
\cal E &= - \frac{g_{tt}}{\sqrt{-g_{tt}-g_{\phi \phi} \Omega^2_{\phi}}}\,, \\
\cal L &=  \frac{g_{tt} \Omega_{\phi}}{\sqrt{-g_{tt}-g_{\phi \phi} \Omega^2_{\phi}}}\,,
\end{align}
\end{subequations}
where the signs ``$\pm$'' are for co- and counter-rotating orbits, respectively \cite{2016EL....11630006B}.

\subsection{The ansatz of small oscillations}

In the regime of small oscillations, the displacements
from equilibrium positions, viz. $ r \sim r_0 + \delta r $ and $\theta \sim \pi/2 + \delta \theta $, imply
\begin{equation}
\frac{d^2 \delta r}{d t} + \Omega^2_{r} \delta r = 0\,, \qquad
\frac{d^2 \delta \theta}{d t} + \Omega^2_{\theta} \delta \theta = 0,
\end{equation}
with corresponding frequencies
\begin{subequations}\label{omegas}
    \begin{align}
    \Omega^2_{r} &=\left. - \frac{1}{2 g_{rr} (u^t)^2} \partial^2_{r} V_{eff}(r,\theta) \right| _{\theta=\pi/2}\,,\\
    \Omega^2_{\theta} &=\left. - \frac{1}{2 g_{\theta \theta} (u^t)^2} \partial^2_{\theta} V_{eff}(r,\theta) \right| _{\theta=\pi/2}\,,
\end{align}
\end{subequations}
for radial and angular coordinates, respectively.
From the above angular frequencies, we define the Keplerian frequency $f_\phi = \Omega _\phi/(2 \pi) $  and the radial epicyclic frequency of the Keplerian motion $f_r = \Omega _r/(2 \pi)$. The relativistic precession model identifies the lower QPO frequency $f_L$ with the periastron precession, namely $f_L=f_\phi - f_r$, and the upper QPO frequency $f_U$ with the Keplerian frequency, namely $f_U = f_\phi$ \cite{2014MNRAS.439L..65M,2014MNRAS.437.2554M}.

\section{Static spherically symmetric regular black holes}

The simplest approach that permits to describe compact object involves the use of a spherically symmetric, non-rotating metric. In particular, spacetime geometry can be characterized through
\begin{equation}
ds^2 = -f(r)dt^2 + f(r)^{-1}dr^2 + r^2(d\theta^2 + \sin^2\theta d\phi^2)\,,
\end{equation}
also involving the class of metrics that exhibit electrically charged object in GR, see e.g. \cite{Muller:2009bw} and furthermore being adaptable to find out solution in extended theories of gravity. Hereafter, we focus on the above class of metric employing regular configurations represented by three RBH solutions: the Bardeen, Hayward and Dymnikova spacetimes. We focused on such metrics since their structures appear particularly simple and apparently well-adaptable to the cases of our interest from which we argue our QPO data points. Below, we elucidate the main features of each spacetime, emphasizing the physical properties that will be discussed in comparing our numerical findings.

\subsection{Bardeen metric}

The Bardeen metric \cite{bardeen1968proceedings,2000PhLB..493..149A} is a solution of Einstein's field equations that represents a non-rotating BH with topological charge, whose lapse function was first derived by John Bardeen in 1968, with the underlying motivation to find a solution to Einstein-Maxwell equations describing  a magnetically-charged BH, as an alternative to the traditional Reissner-Nordstr\"{o}m BH solution.

Consequently, the lapse function can be written in the following form
\begin{equation}
f (r) =  1 - \frac{2 M r^2}{(r^2 + q^2)^{3/2}}\,,
\end{equation}
where  $q$ and $M$ are the magnetic charge and the mass of the magnetic monopole, respectively. For vanishing $q$ the Bardeen metric reduces to the Schwarzschild black hole.

This spacetime solution could be somehow reinterpreted as the prototype of \emph{quasi-Kerr solution}, where
the metric tensor in Boyer-Lindquist coordinates shows the same expression of Kerr metric, but replacing the mass through a mass function depending on the radial coordinate $r$, with the physical property to become the standard mass as $r$ approaches large distances\footnote{This represents a class of metrics. This class degenerates with Kerr and provides extremely similar results to the Kerr BH.} \cite{Bambi:2014nta}.

Thus, the Bardeen metric appears particularly appealing to test compact object QPOs and then, adopting the same strategy presented in Eqs. \eqref{omegas}, we can compare our findings for the Bardeen spacetime with numerical data. By adopting the Bardeen metric, one can write the fundamental frequencies for test particles
\begin{subequations}
\begin{align}
\Omega^2 _\phi &= \frac{M \left(r^2-2 q^2\right)}{\left(r^2+q^2\right)^{5/2}}\,, \\
\Omega^2 _\theta &= \Omega^2 _\phi\,, \\
\Omega^2 _r &=\frac{M\left(r^6+9 q^2 r^4-8 q^6\right)}{\left(r^2+q^2\right)^{9/2}}-\frac{6 M^2 r^6}{\left(r^2+q^2\right)^5}
\,,
\end{align}
\end{subequations}
where $\Omega _\phi $ represents the Keplerian angular velocity of a test particle measured by an observer placed at infinity, $\Omega _r $ is the radial angular velocity and $\Omega _\theta $ is the vertical angular velocity. Last but not least, it is remarkable to stress that the physical interpretation of $q$ is provided by the existence of monopole charge of the self-gravitating magnetic field. Thus, the Bardeen metric, if applied to exteriors of compact objects, will provide information about the net charge associated to them, in terms of a monopole nonlinear charge. We will check later the goodness of this hypothesis in the framework of QPOs.

\subsection{Hayward metric}

In 2006, Hayward  found a new regular BH spacetime that resembles the physical interpretation of the Bardeen one and has center flatness. This simple RBH implies a specific matter energy-momentum tensor that is de Sitter at the core and vanishes at large distances $r \to \infty$ \cite{2006PhRvL..96c1103H}. The lapse function $f(r)$ for the Hayward BH takes a simple form
\begin{equation}
f (r) =  1 - \frac{2 M r^2}{r^3 + 2 a^2}\,,
\end{equation}
where $M$ is the mass of the BH, $r$ is the radial coordinate, $a$ is a constant.

In analogy to what we have prompted for the Bardeen spacetime, one can write fundamental frequencies as follows
\begin{subequations}
\begin{align}
\Omega^2 _\phi &= \frac{M \left(r^3-4 a^2\right)}{\left(r^3+2 a^2\right)^2}\,, \\
\Omega^2 _\theta &= \Omega^2 _\phi\,, \\
\Omega^2 _r &= \frac{M \left[r^5 (r-6 M)+22 a^2 r^3-32 a^4\right]}{\left(r^3+2 a^2\right)^3}\,,
\end{align}
\end{subequations}
where, as before, $\Omega _\phi $ represents the Keplerian angular velocity of a test particle measured by an observer placed at infinity, $\Omega _r $ is the radial angular velocity and $\Omega _\theta $ is the vertical angular velocity. For the sake of completeness, it is remarkable to notice that both Bardeen and Hayward metrics can be  inspired by the Damour–Solodukhin scenario where a line element, with distinct lapse and shift functions, has been introduced to furnish class of solutions among which Bardeen and Hayward \cite{DuttaRoy:2022ytr}.

\begin{figure*}
{\hfill
\includegraphics[width=0.41\hsize,clip]{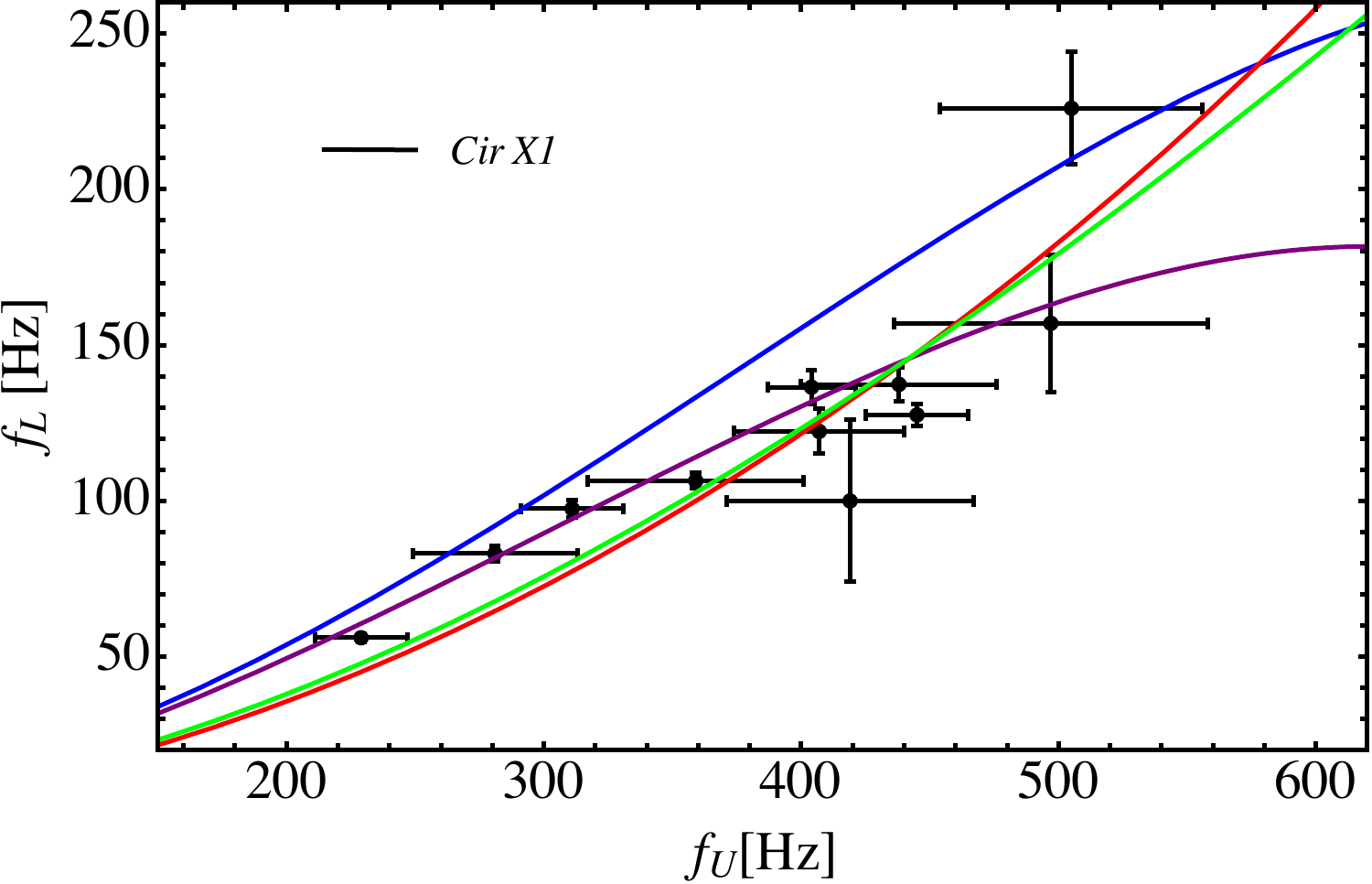}\hfill
\includegraphics[width=0.41\hsize,clip]{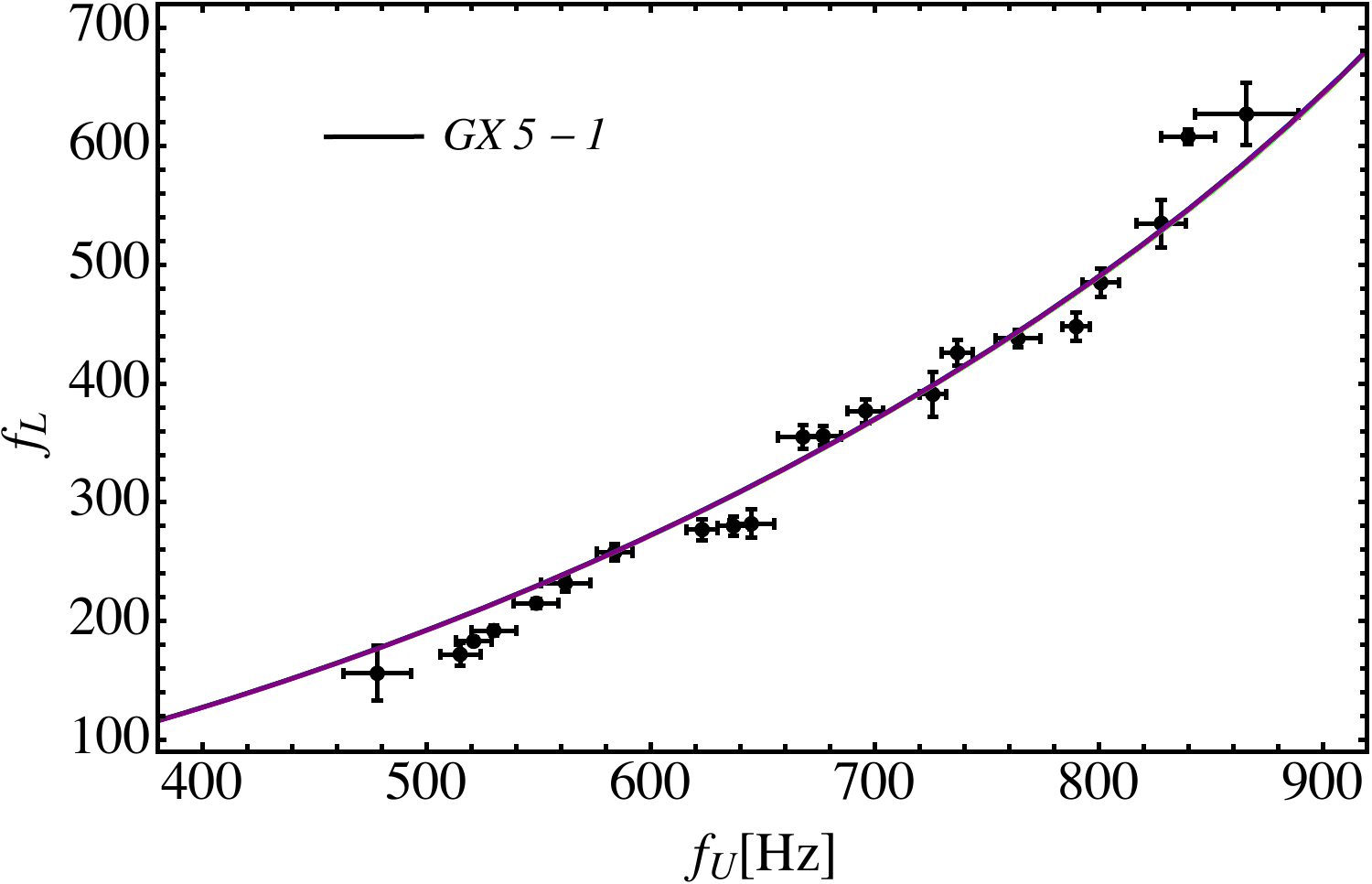}\hfill}

{\hfill
\includegraphics[width=0.41\hsize,clip]{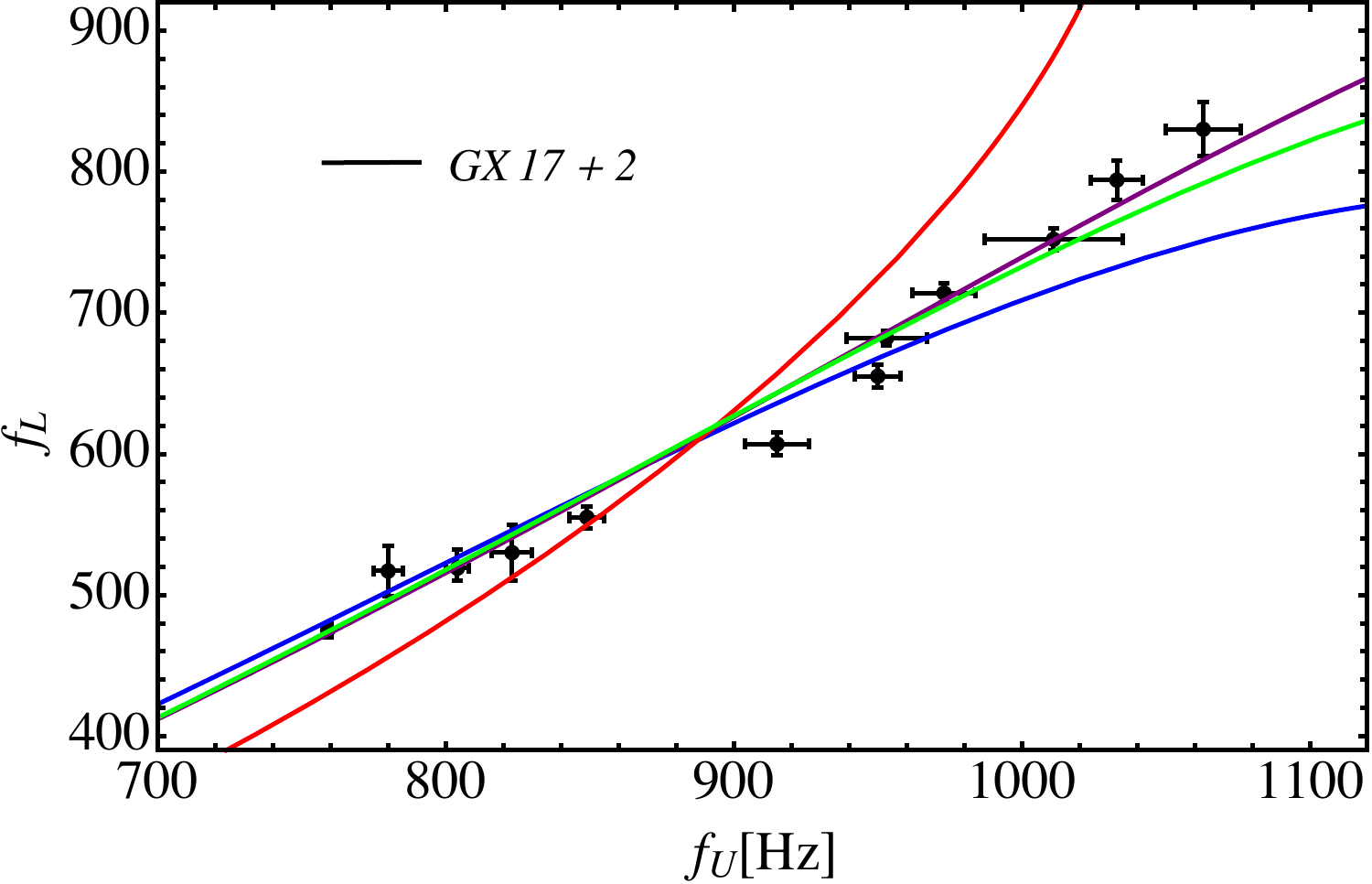}\hfill
\includegraphics[width=0.41\hsize,clip]{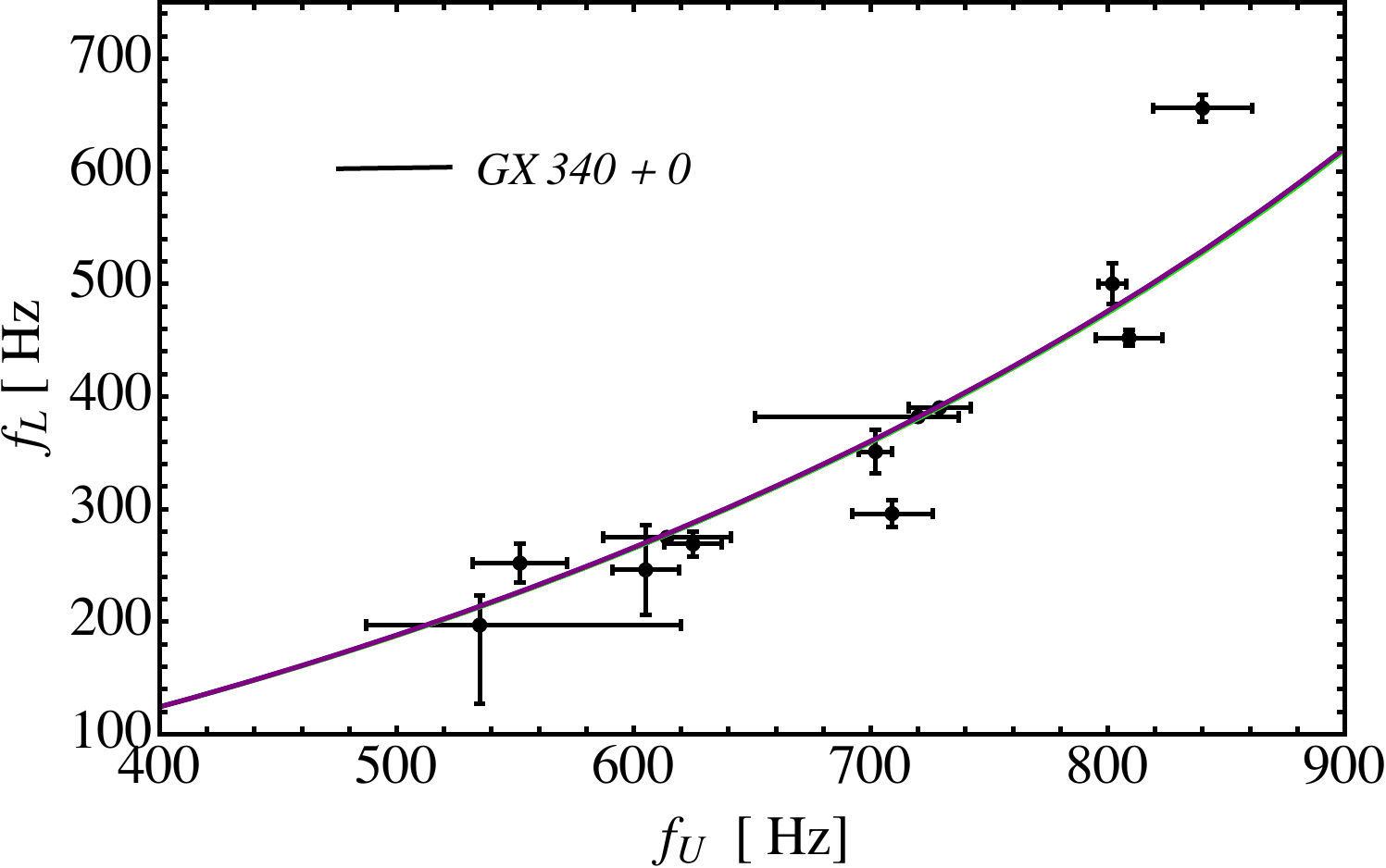}\hfill}

{\hfill
\includegraphics[width=0.41\hsize,clip]{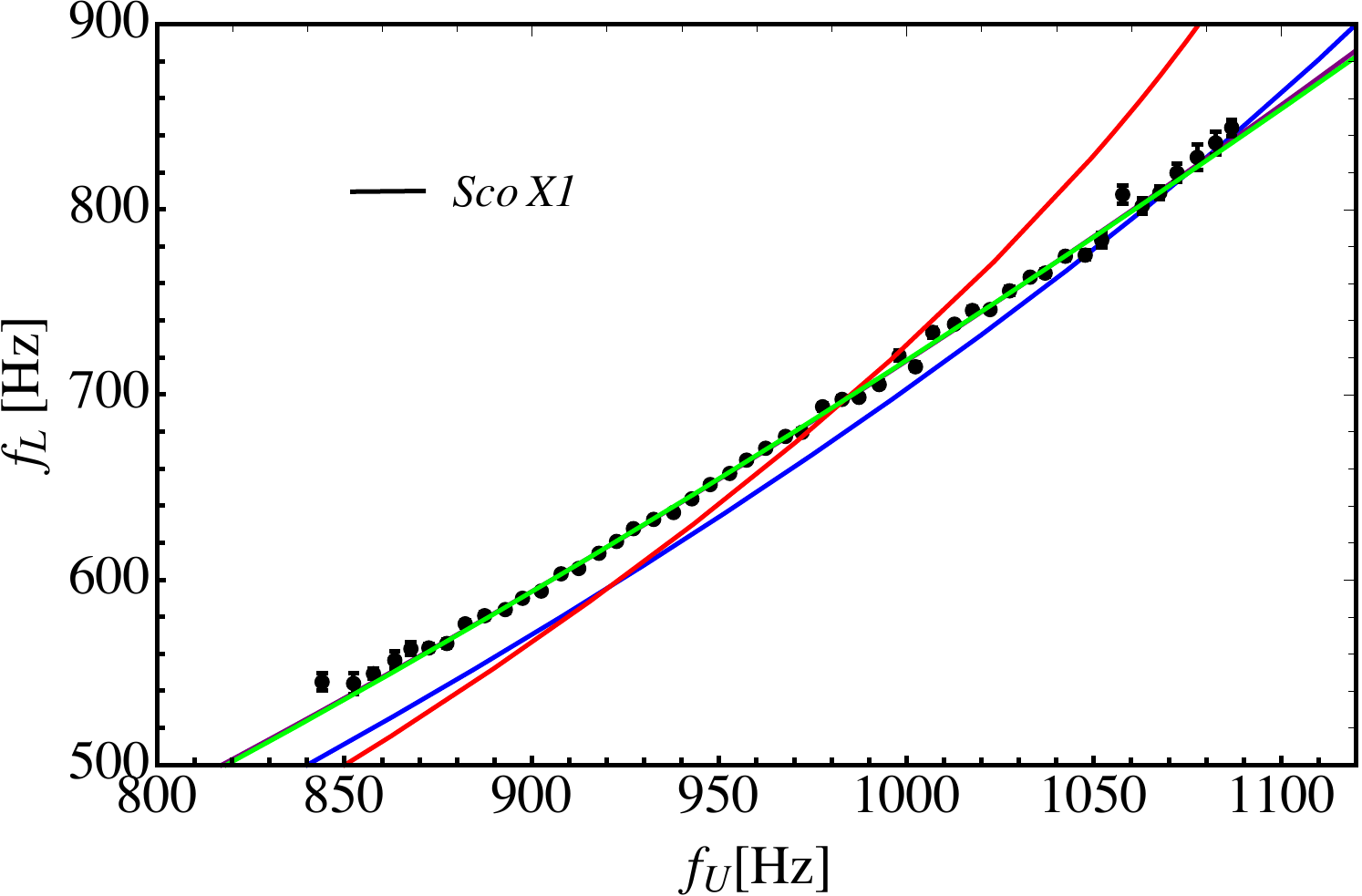}\hfill
\includegraphics[width=0.41\hsize,clip]{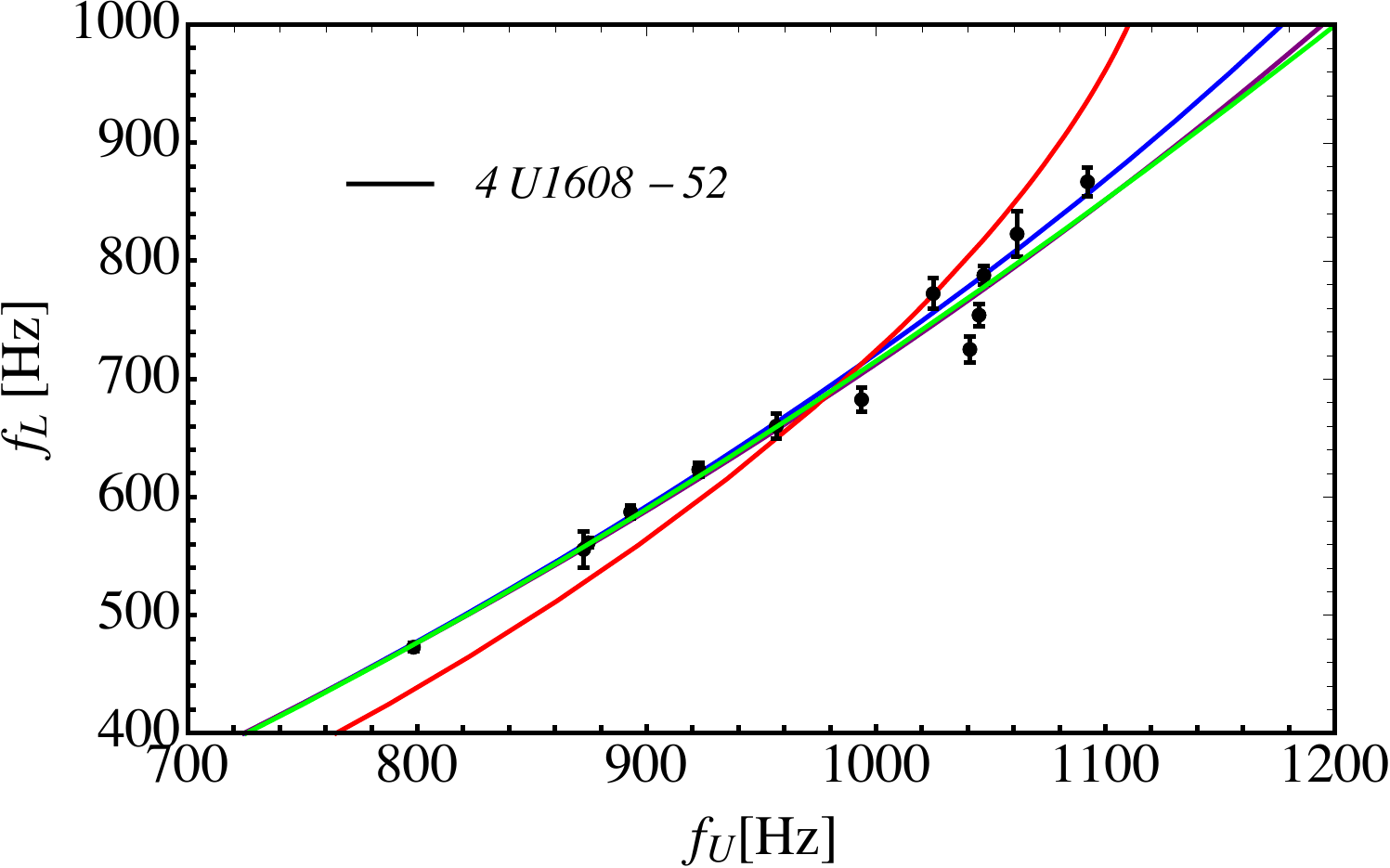}\hfill}

{\hfill
\includegraphics[width=0.41\hsize,clip]{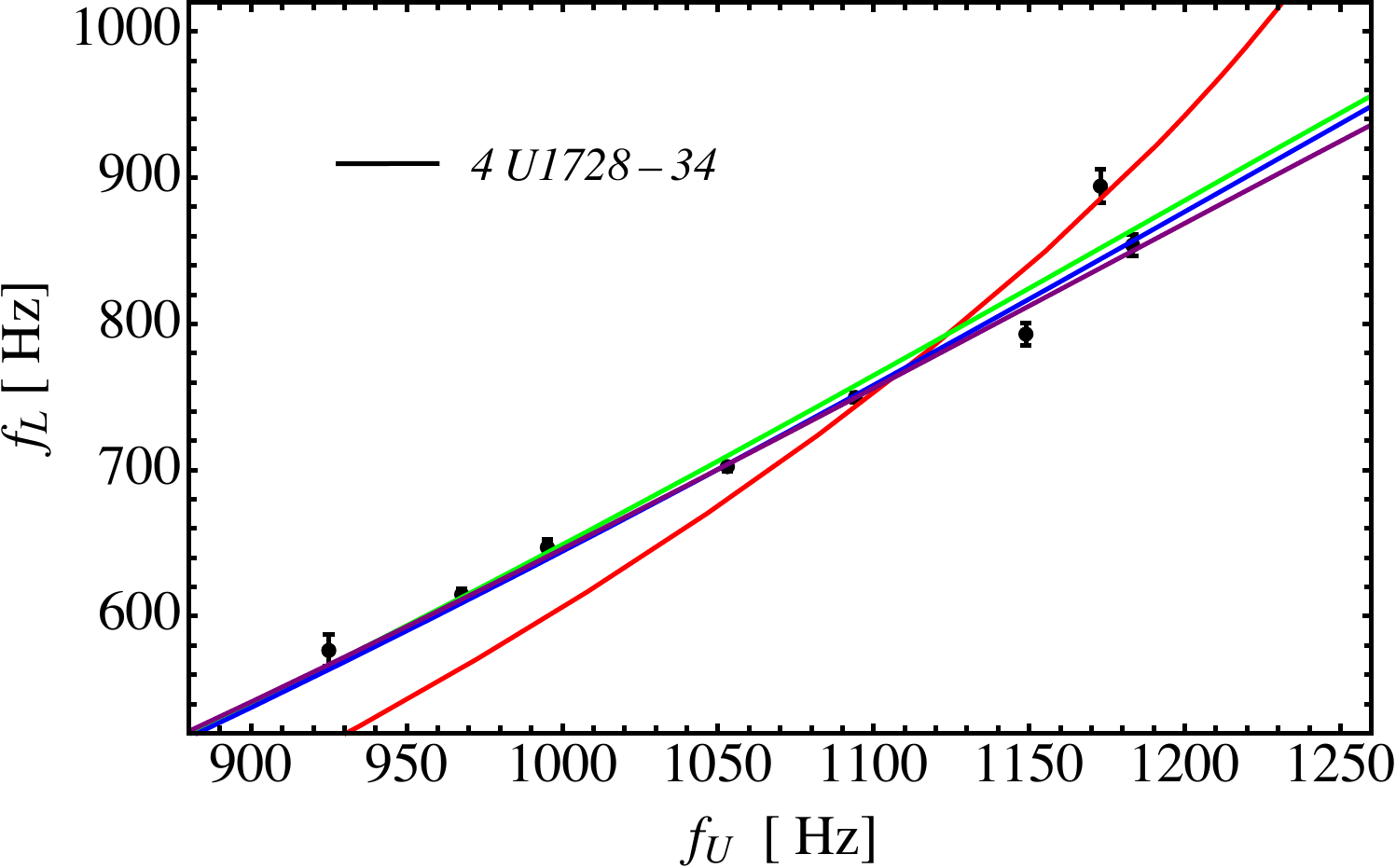}\hfill
\includegraphics[width=0.41\hsize,clip]{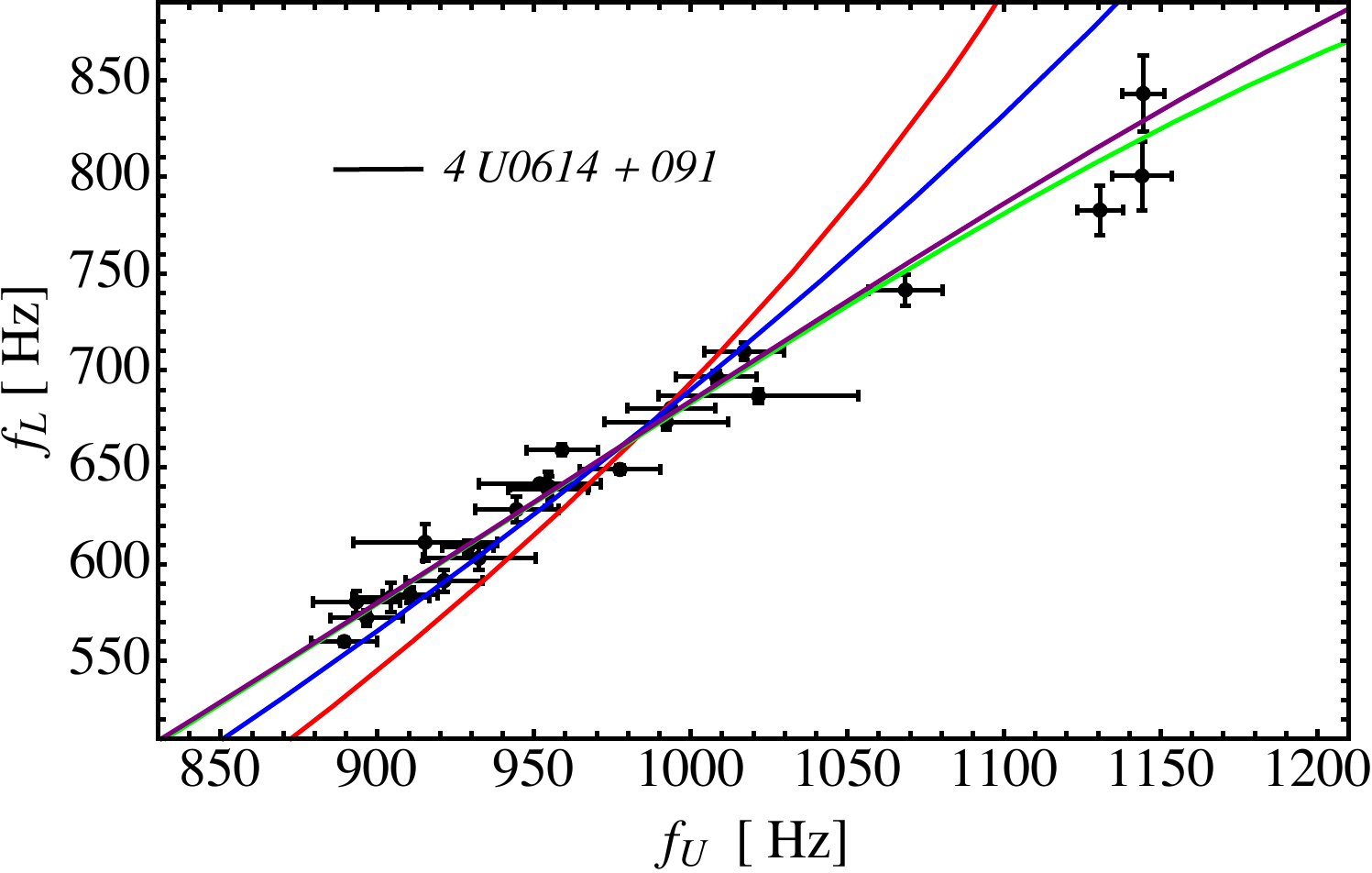}\hfill}
\caption{Plots of $f_{\rm L}$ vs $f_{\rm U}$ frequencies of the QPO data sets considered in this work (black data with error bars). Fits have been performed by using  Schwarzschild (red), Hayward (green), Bardeen (purple) and Dymnikova (blue) spacetimes.}
\label{fig:freq}
\end{figure*}

\subsection{Dymnikova metric}

The Dymnikova metric is based on the idea that the gravitational field of a BH can be described by a non-symmetric metric, which takes into account the presence of a non-zero energy-momentum tensor in proximity of the BH \cite{2004CQGra..21.4417D}. This approach allows for a more accurate description of the gravitational field of a BH and its effects on the surrounding space-time. The lapse function for Dymnikova's solution is of the following form
\begin{equation}
    f(r) = 1-\frac{2 M}{r} \frac{2}{\pi} \left[\arctan \left(\frac{r}{r_0}\right)
-\frac{rr_0}{r^2+r_0^2}\right]\,,
\end{equation}
where $r_0$ is the length scale, $r_0=\pi q^2/(8 M)$, $M$ is the total mass, $q$ is the charge.
By adopting the Dymnikova metric, one can calculate all fundamental angular frequencies
\begin{widetext}
\begin{subequations}
\begin{align}
\Omega^2 _\phi &= \frac{2 M}{\pi r^3} \left[\arctan \left(\frac{r}{r_0}\right) - \frac{r r_0(3 r^2+r_0^2)}{(r^2 +
   r_0^2)^2} \right]\,,\\
\Omega^2 _\theta &= \Omega^2 _\phi\,, \\
\Omega^2 _r & = \frac{2 M}{\pi^2 r^4 (r^2 + r_0^2)^3} \Bigg\{ r^2 r_0 \left[-4 M r_0 \left(17 r^2 + 3 r_0^2\right)+\pi  \left(r^4-8 r^2 r_0^2-r_0^4\right)\right]+  \\ \nonumber
&  \ \ \ \arctan\left(\frac{r}{r_0}\right) \left[-12 M \left(r^2 + r_0^2\right)^3 \arctan\left(\frac{r}{r_0}\right) +8 M r r_0 \left(6 r^4+13 r^2 r_0^2+3 r_0^4\right)+\pi  r \left(r^2 +r_0^2\right)^3\right]\Bigg\}\,.
\end{align}
\end{subequations}
\end{widetext}
\begin{table*}
\centering
\setlength{\tabcolsep}{.5em}
\renewcommand{\arraystretch}{1.3}
\begin{tabular}{lccccrrrrr}
\hline\hline
Source                                  &
Metric                                  &
$M$                                     &
$a$                                     &
$q$                                     &
$\ln L$                                 &
AIC                                     &
BIC                                     &
$\Delta$AIC                             &
$\Delta$BIC                             \\
                                        &
                                        &
$(M_\odot)$                             &
$({\rm km}^{3/2})$                      &
$({\rm km})$                            & & & & &\\
\hline
Cir X1                                  &
S                                       &
$2.224^{+0.029}_{-0.029}$               &
--                                      &
--                                      &
$-125.84$                               &
$254$ & $254$ & $145$ & $145$ \\
                                        &
H                                       &
$2.466^{+0.061}_{-0.038}$               &
$9.928^{+0.071}_{-0.612}$               &
--                                      &
$-105.77$                               &
$216$ & $216$ & $107$ & $107$ \\
                                        &
B                                       &
$4.74^{+0.23}_{-0.23}$                  &
--                                      &
$9.34^{+0.47}_{-0.52}$                  &
$-52.23$                                &
$109$ & $109$ & $0$ & $0$ \\
                                        &
D                                       &
$7.49^{+0.07}_{-0.84}$                  &
--                                      &
$14.78^{+0.13}_{-1.71}$                 &
$-52.33$                                &
$109$ & $109$ & $0$ & $0$ \\
\hline
GX 5--1                                 &
S                                       &
$2.161^{+0.010}_{-0.010}$               &
--                                      &
--                                      &
$-200.33$ &
$403$ & $404$ & $0$ & $0$ \\
                                        &
H                                       &
$2.163^{+0.015}_{-0.017}$               &
$0.30^{+0.63}_{-0.30}$                  &
--                                      &
$-200.32$                               &
$405$ & $407$ & $2$ & $3$ \\
                                        &
B                                       &
$2.164^{+0.017}_{-0.020}$               &
--                                      &
$0.14^{+0.28}_{-0.14}$                  &
$-200.34$                               &
$405$ & $407$ & $2$ & $3$ \\
                                        &
D                                       &
$2.177^{+0.046}_{-0.032}$               &
--                                      &
$0.47^{+0.53}_{-0.46}$                  &
$-200.44$                               &
$405$ & $407$ & $2$ & $3$ \\
\hline
GX 17+2                                 &
S                                       &
$2.07678^{+0.0002}_{-0.0003}$           &
--                                      &
--                                      &
$-1819.02$ &
$3640$ & $3641$ & $3513$ & $3513$ \\
                                        &
H                                       &
$3.111^{+0.023}_{-0.022}$               &
$12.80^{+0.16}_{-0.22}$                 &
--                                      &
$-61.67$                                &
$127$ & $128$ & $0$ & $0$ \\
                                        &
B                                       &
$3.803^{+0.008}_{-0.020}$               &
--                                      &
$5.488^{+0.009}_{-0.043}$               &
$-61.81$                                &
$128$ & $129$ & $1$ & $1$ \\
                                        &
D                                       &
$5.8124^{+0.0064}_{-0.0073}$            &
--                                      &
$10.180^{+0.014}_{-0.012}$              &
$-217.15$                               &
$438$ & $439$ & $311$ & $311$ \\
\hline
GX 340+0                                &
S                                       &
$2.102^{+0.003}_{-0.003}$               &
--                                      &
--                                      &
$-130.86$ &
$264$ & $264$ & $0$ & $0$ \\
                                        &
H                                       &
$2.112^{+0.013}_{-0.008}$               &
$0.60^{+1.02}_{-0.59}$                  &
--                                      &
$-134.17$                               &
$273$ & $273$ & $9$ & $9$ \\
                                        &
B                                       &           $2.113^{+0.025}_{-0.010}$               &
--                                      &
$0.28^{+0.42}_{-0.26}$                  &
$-134.13$                               &
$273$ & $273$ & $9$ & $9$ \\
                                        &
D                                       &
$2.137^{+0.081}_{-0.036}$               &
--                                      &
$0.68^{+0.67}_{-0.62}$                  &
$-134.10$                               &
$273$ & $273$ & $9$ & $9$ \\
\hline
Sco X1                                  &
S                                       &
$1.9649^{+0.0005}_{-0.0005}$            &
--                                      &
--                                      &
$-3887.17$                              &
$7776$ & $7779$ & $7503$ & $7503$ \\
                                        &
H                                       &
$2.659^{+0.014}_{-0.006}$               &
$9.573^{+0.116}_{-0.050}$               &
--                                      &
$-137.03$                               &
$278$ & $281$ & $5$ & $5$ \\
                                        &
B                                       &
$3.242^{+0.017}_{-0.015}$               &
--                                      &
$4.536^{+0.035}_{-0.029}$               &
$-134.56$                               &
$273$ & $276$ & $0$ & $0$ \\
                                        &
D                                       &
$3.83^{+0.04}_{-0.71}$                 &
--                                      &
$6.37^{+0.05}_{-1.47}$                  &
$-2591.17$                               &
$5187$ & $5190$ & $4914$ & $4914$  \\
\hline
4U1608--52                              &
S                                       &
$1.960^{+0.004}_{-0.004}$               &
--                                      &
--                                      &
$-235.83$   &
$474$ & $475$ & $348$ & $348$ \\
                                        &
H                                       &                       $2.627^{+0.044}_{-0.039}$               &
$9.35^{+0.37}_{-0.32}$                  &
--                                      &
$-60.96$                                &
$127$ & $128$ & $1$ & $1$ \\
                                        &
B                                       &
$3.180^{+0.061}_{-0.064}$               &
--                                      &
$4.43^{+0.12}_{-0.13}$                  &
$-60.85$                                &
$126$ & $127$ & $0$ & $0$ \\
                                        &
D                                       &
$4.37^{+0.02}_{-0.12}$                  &
--                                      &
$7.44^{+0.04}_{-0.24}$                  &
$-65.35$                                &
$135$ & $136$ & $9$ & $9$ \\
\hline
4U1728--34                              &
S                                       &
$1.734^{+0.003}_{-0.003}$               &
--                                      &
--                                      &
$-212.61$   &
$427$ & $427$ & $349$ & $349$ \\
                                        &
H                                       &
$2.470^{+0.034}_{-0.046}$               &
$8.98^{+0.25}_{-0.34}$                  &
--                                      &
$-45.17$                                &
$94$ & $94$ & $15$ & $15$ \\
                                        &
B                                       &
$3.055^{+0.062}_{-0.064}$               &
--                                      &
$4.41^{+0.11}_{-0.12}$                  &
$-37.27$                                &
$79$ & $79$ & $0$ & $0$ \\
                                        &
D                                       &
$4.170^{+0.021}_{-0.014}$               &
--                                      &
$7.215^{+0.041}_{-0.028}$               &
$-38.47$                                &
$81$ & $81$ & $2$ & $2$ \\
\hline
4U0614+091                              &
S                                       &
$1.904^{+0.001}_{-0.001}$               &
--                                      &
--                                      &
$-842.97$   &
$1688$ & $1691$ & $1398$ & $1398$ \\
                                        &
H                                       &
$2.808^{+0.028}_{-0.022}$               &
$11.11^{+0.21}_{-0.16}$                 &
--                                      &
$-143.09$                                &
$290$ & $293$ & $0$ & $0$ \\
                                        &
B                                       &
$3.509^{+0.011}_{-0.032}$               &
--                                      &
$5.126^{+0.023}_{-0.054}$               &
$-146.22$                                &
$296$ & $299$ & $6$ & $6$  \\
                                        &
D                                       &
$4.0282^{+0.0041}_{-0.0079}$            &
--                                      &
$6.813^{+0.008}_{-0.017}$               &
$-266.85$                                &
$538$ & $541$ & $248$ & $248$  \\
\hline
\end{tabular}
\caption{Best-fit parameters with the associated $1$--$\sigma$ error bars obtained from Hayward (H), Bardeen (B) and Dymnikova (D) metrics. $\Delta$AIC and $\Delta$BIC are computed with respect to the reference model, \textit{i.e.}, the model with the highest value of $\ln L$. For comparisons, we also reported the results from the Schwarzschild (S) metric obtained in Ref.~\cite{2022arXiv221210186B}.}
\label{tab:results}
\end{table*}

\begin{table}
\centering
\setlength{\tabcolsep}{1.em}
\renewcommand{\arraystretch}{1.1}
\begin{tabular}{lcccc}
\hline
\hline
Source & Model & ISCO &  Inner &  Outer \\
       &       & (km) &  (km)  &  (km) \\
\hline
Cir X-1 & S & $19.62$ & $30.79$    &  $52.16 $  \\
$\,$ &  H & 5.74  & $31.60$ & $53.90$\\
$\,$ & B & 9.76 & $35.70$ & $65.08$\\
$\,$ & D & 10.44 & $36.24$ & $70.13$\\
\hline
GX 5-1 & S & $19.06$    &  $21.29$ & $31.63$   \\
$\,$ &  H & $19.07$  & $21.29$ & $31.64$\\
$\,$ & B & $19.08$  & $21.29$ & $31.65$\\
$\,$ & D & $19.10$  & $21.31$ & $31.69$\\
\hline
GX 17+2 & S & $18.32$    &  $18.33$ & $22.94$   \\
$\,$ &  H &
6.88 & $19.80$ & $25.56$\\
$\,$ & B &
 6.12 & $19.98$ & $26.27$\\
$\,$ & D & 6.80 & $19.33$ & $26.71$\\
\hline
GX 340+0 & S & $18.54$    &  $21.52$ & $29.07$   \\
$\,$ &  H & $18.61$  & $21.55$ & $29.12$\\
$\,$ & B & $18.60$  & $21.55$ & $29.12$\\
$\,$ & D & $18.62$  & $21.59$ & $29.19$\\
\hline
Sco X1 & S & $17.33$ &  $17.72$ & $20.98$   \\
$\,$ &  H & 5.72 & $18.90$ & $22.72$\\
$\,$ & B &   16.02 & $19.25$ & $23.42$\\
$\,$ & D & 17.08 & $19.00$ & $23.22$\\
\hline
4U1608-52 & S & $17.29$ &  $17.65$ & $21.75$   \\
$\,$ &  H & 5.64 & $18.78$ & $23.56$\\
$\,$ & B &
16.22 & $19.12$ & $24.31$\\
$\,$ & D & 16.58 & $19.08$ & $24.74$\\
\hline
4U1728-34 & S & $15.29$ &  $16.06$ & $18.93$   \\
$\,$ &  H & 5.44 & $17.33$ & $20.79$\\
$\,$ & B & 4.92 & $17.64$ & $21.45$\\
$\,$ & D & 4.81 & $17.49$ & $21.62$\\
\hline
4U0614+091 & S & $16.80$ &  $16.95$ & $20.05$   \\
$\,$ &  H & 6.26 & $18.26$ & $22.13$\\
$\,$ & B & 5.70& $18.47$ & $22.76$\\
$\,$ & D & 16.26 & $18.19$ & $22.39$\\
\hline
\end{tabular}
\caption{Numerical values of ISCO and inner and outer disk radii for each source have been computed from best-fit results of Table~\ref{tab:results} for Schwarzschild (S), Hayward (H), Bardeen (B) and Dymnikova (D) metrics. Thus, according to the ISCO values one can not exclude/rule out RBH solutions.
}
\label{tab:isco}
\end{table}

\section{Numerical analysis}

In order to numerically check our theoretical frameworks with data, we modify the free-available \texttt{Wolfram Mathematica} code developed in Ref.~\cite{2019PhRvD..99d3516A} to work out our  MCMC analyses, based on the Metropolis-Hastings algorithm. We thus find the best-fit parameters resulting from the maximum of the log-likelihood, defined by
\begin{equation}
\label{loglike}
    \ln L = -\sum_{k=1}^{N}\left\{\dfrac{\left[f_{\rm L}^k-f_{\rm L}(p,\overline{f_{\rm U}^k})\right]^2}{2(\sigma f_{\rm L}^k)^2} + \ln(\sqrt{2\pi}\sigma f_{\rm L}^k)\right\}\,,
\end{equation}
where $p$ labels the model parameters and $N$ the data for each source, sampled as lower frequencies $f_{\rm L}^k$, attached errors $\sigma f_{\rm L}^k$, and \textit{error-averaged} upper frequencies $\overline{f_{\rm U}^k}$, which will be better explained below.
Not all the metrics considered in this work have analytic expressions for $f_{\rm L} =f_{\rm L}(p,f_{\rm U})$. Therefore, to run our MCMC simulations, we followed the steps described below.
\begin{itemize}
\item[I.] For each data point $f_{\rm U}^k$ and each metric, by solving numerically the equations $f_{\rm U}^k\equiv f_{\rm U}(p,r^k)$, one can find the radial coordinate solutions $r^k\equiv r^k(p,f_{\rm U}^k)$.
\item[II.] To account for the errors $\sigma f_{\rm U}^k$, we find the solutions $r_+^k\equiv r_+^k(p,f_{\rm U}^k+\sigma f_{\rm U}^k)$ and $r_-^k\equiv r_-^k(p,f_{\rm U}^k-\sigma f_{\rm U}^k)$, where ``$+$'' and ``$-$'' indicate the solutions of $r$ obtained from $f_{\rm U}^k+\sigma f_{\rm U}^k$ and $f_{\rm U}^k-\sigma f_{\rm U}^k$, respectively.
\item[III.] We compute \textit{error-averaged} solutions $\overline{r^k}=(r^k_+ + r^k_-)/2$, which essentially are functions of what can be defined as \textit{error-averaged} upper frequencies $\overline{f_{\rm U}^k}$.
\item[IV.] Finally, we compute the theoretical lower frequencies $f_{\rm L} =f_{\rm L}[p,\overline{r^k}(p,\overline{f_{\rm U}^k})]\equiv f_{\rm L}(p,\overline{f_{\rm U}^k})$, which implicitly include also the uncertainties $\sigma f_{\rm U}^k$.
\end{itemize}

\begin{figure*}[t]
{\hfill
\includegraphics[width=0.24\hsize,clip]{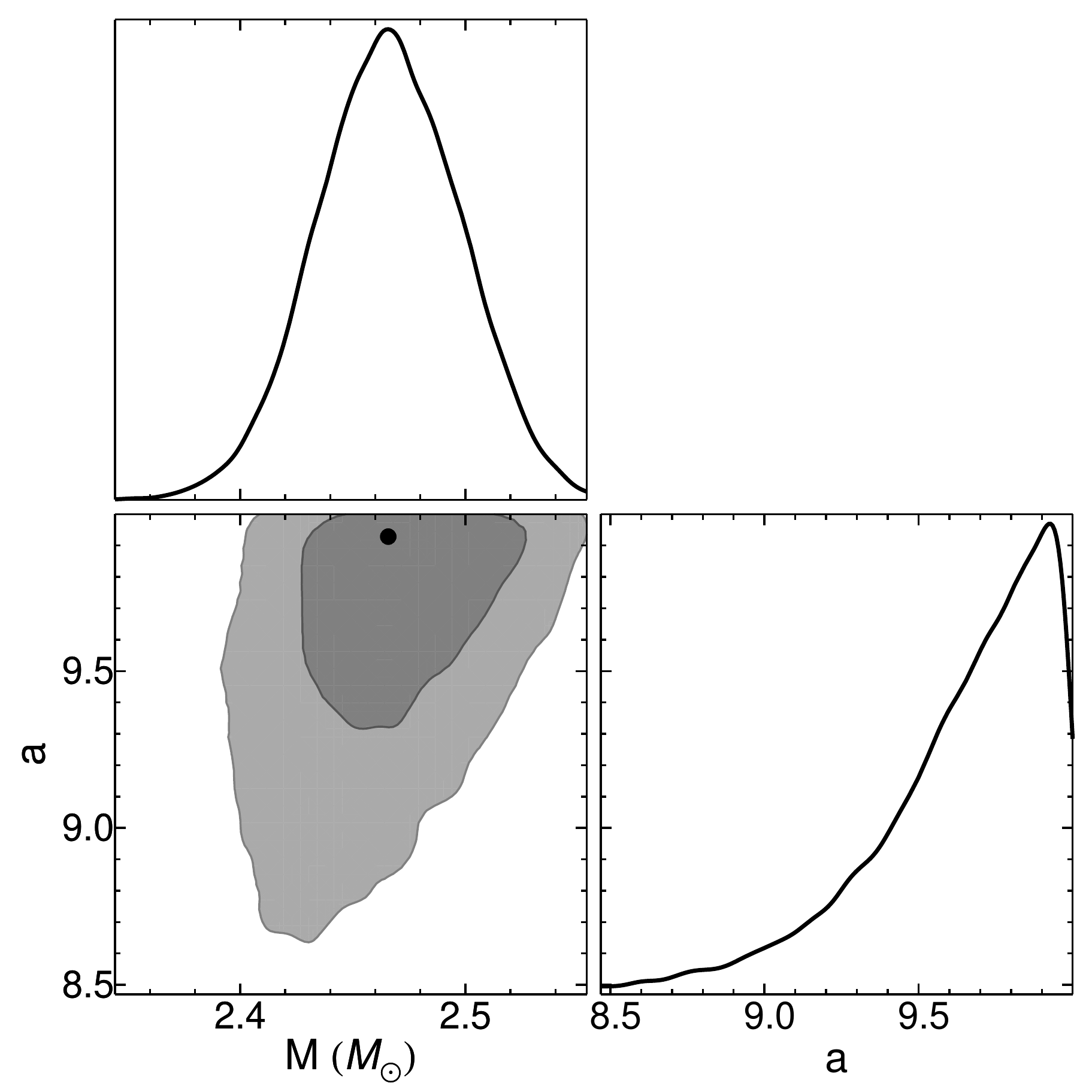}\hfill
\includegraphics[width=0.24\hsize,clip]{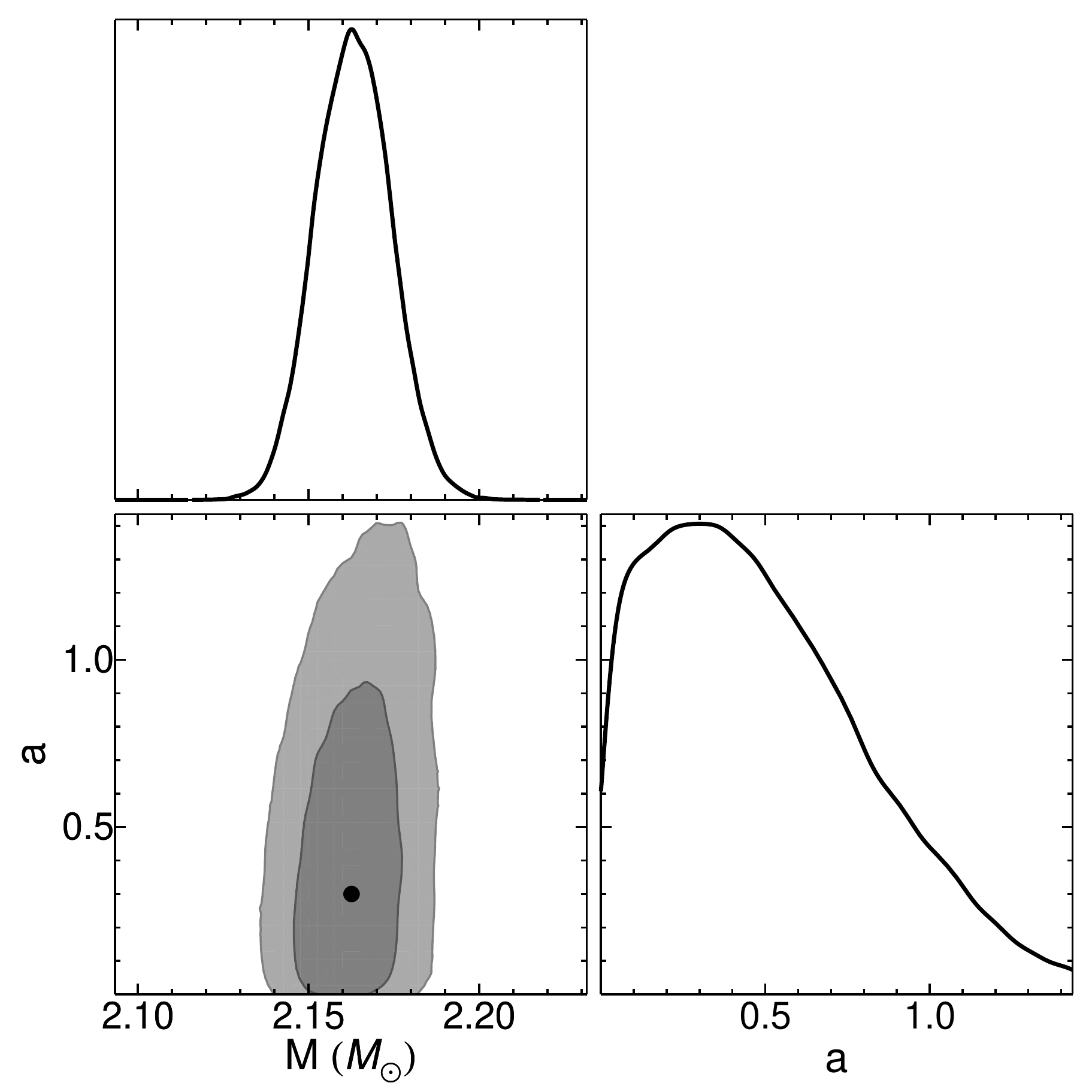}
\hfill
\includegraphics[width=0.24\hsize,clip]{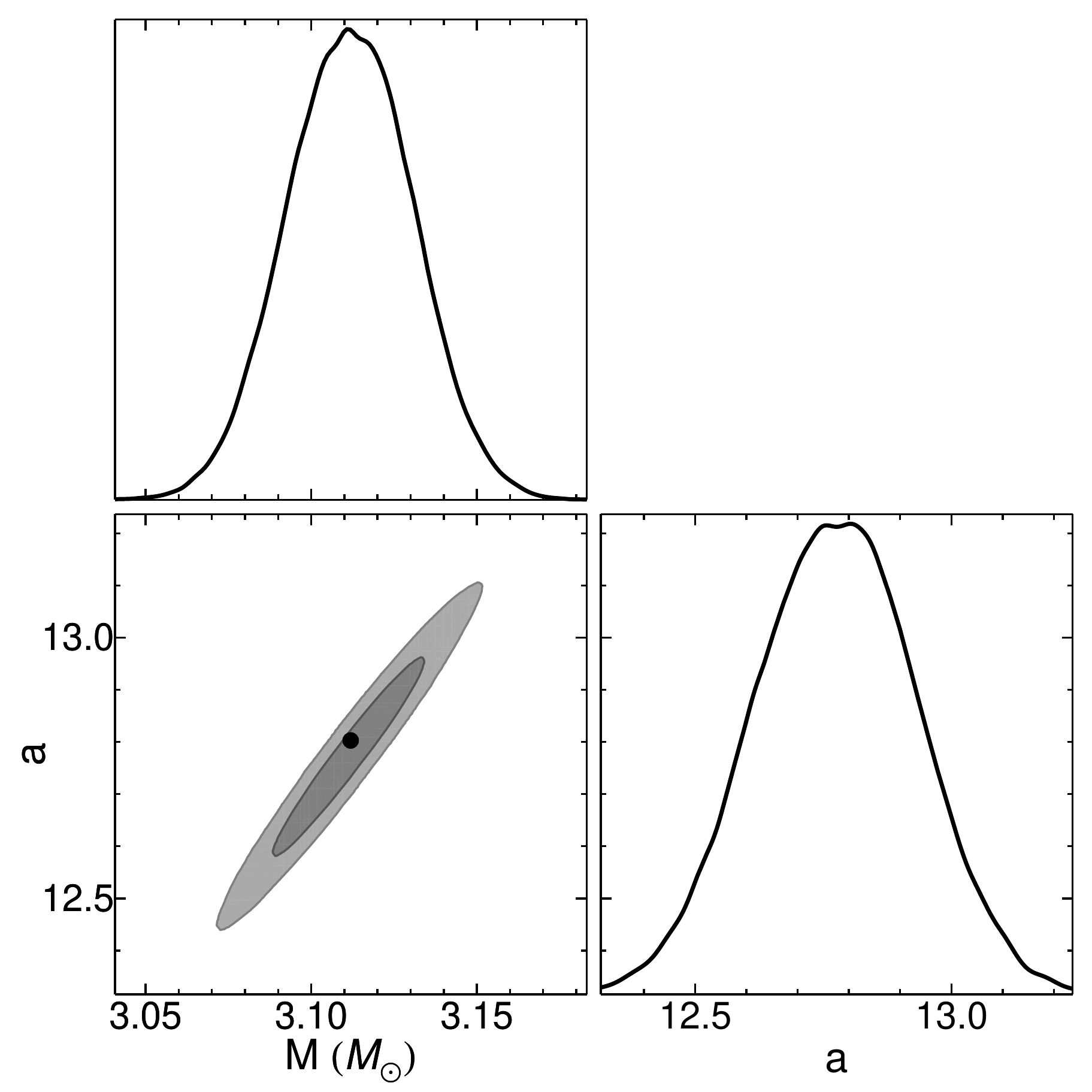}\hfill
\includegraphics[width=0.24\hsize,clip]{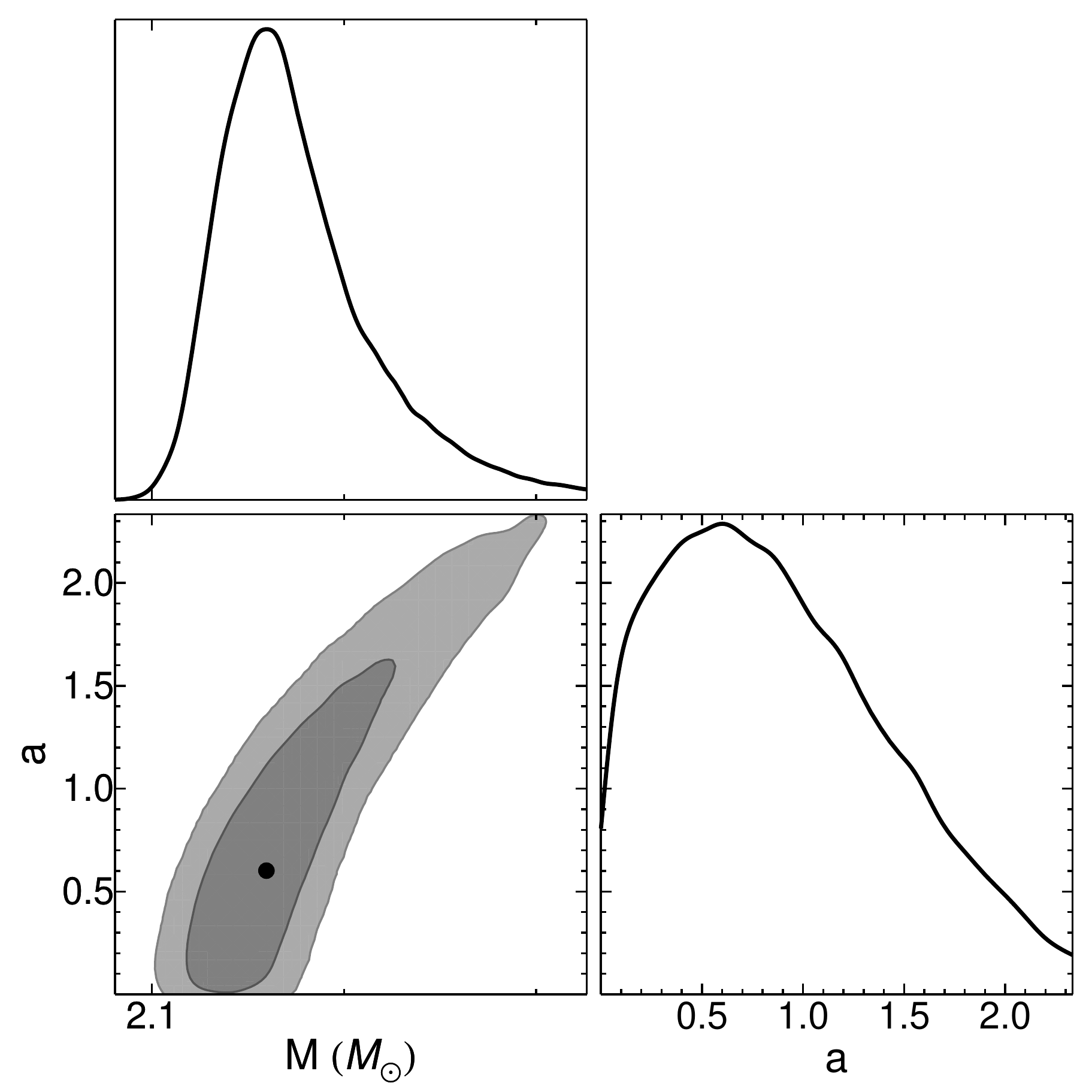}\hfill}

{\hfill
\includegraphics[width=0.24\hsize,clip]{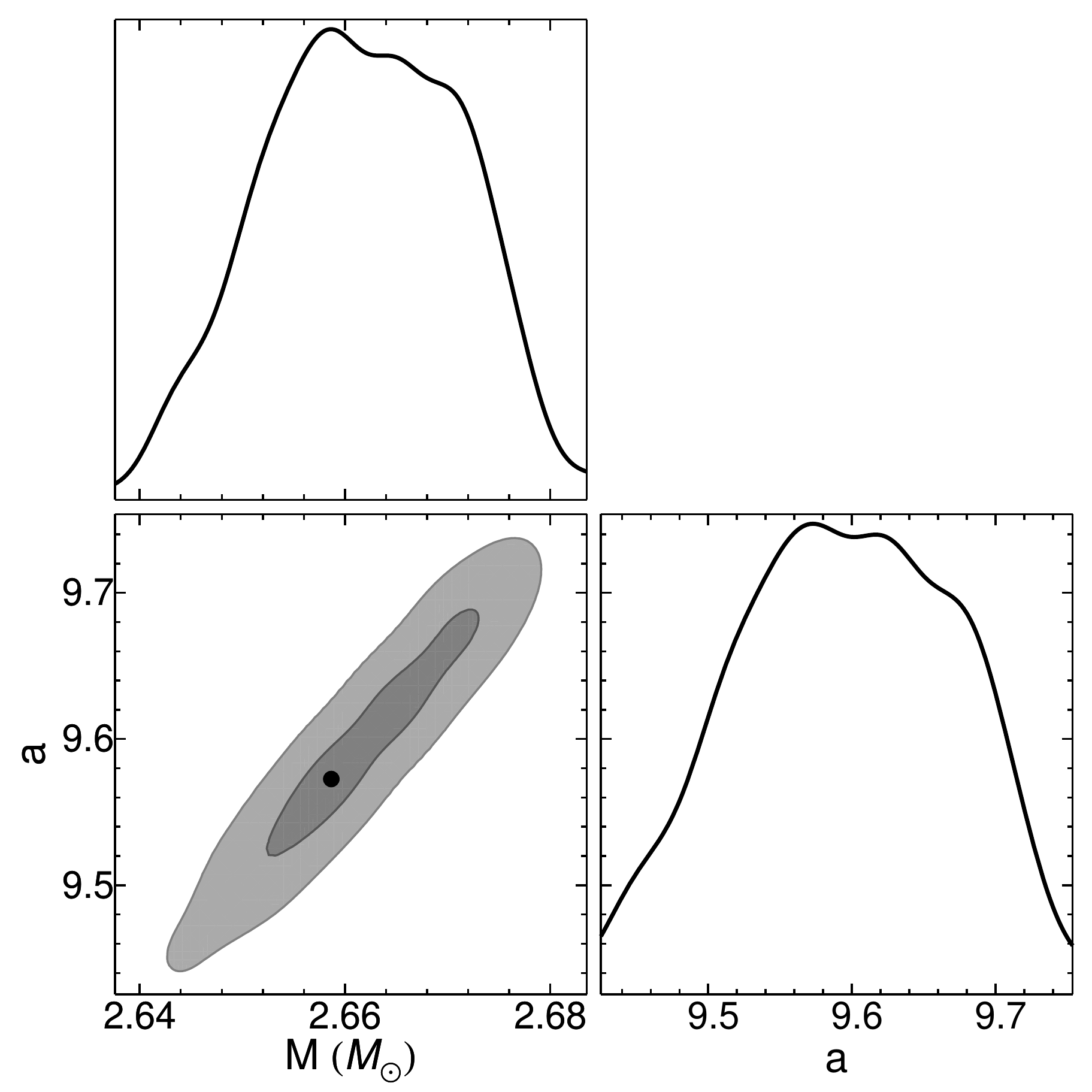}\hfill
\includegraphics[width=0.24\hsize,clip]{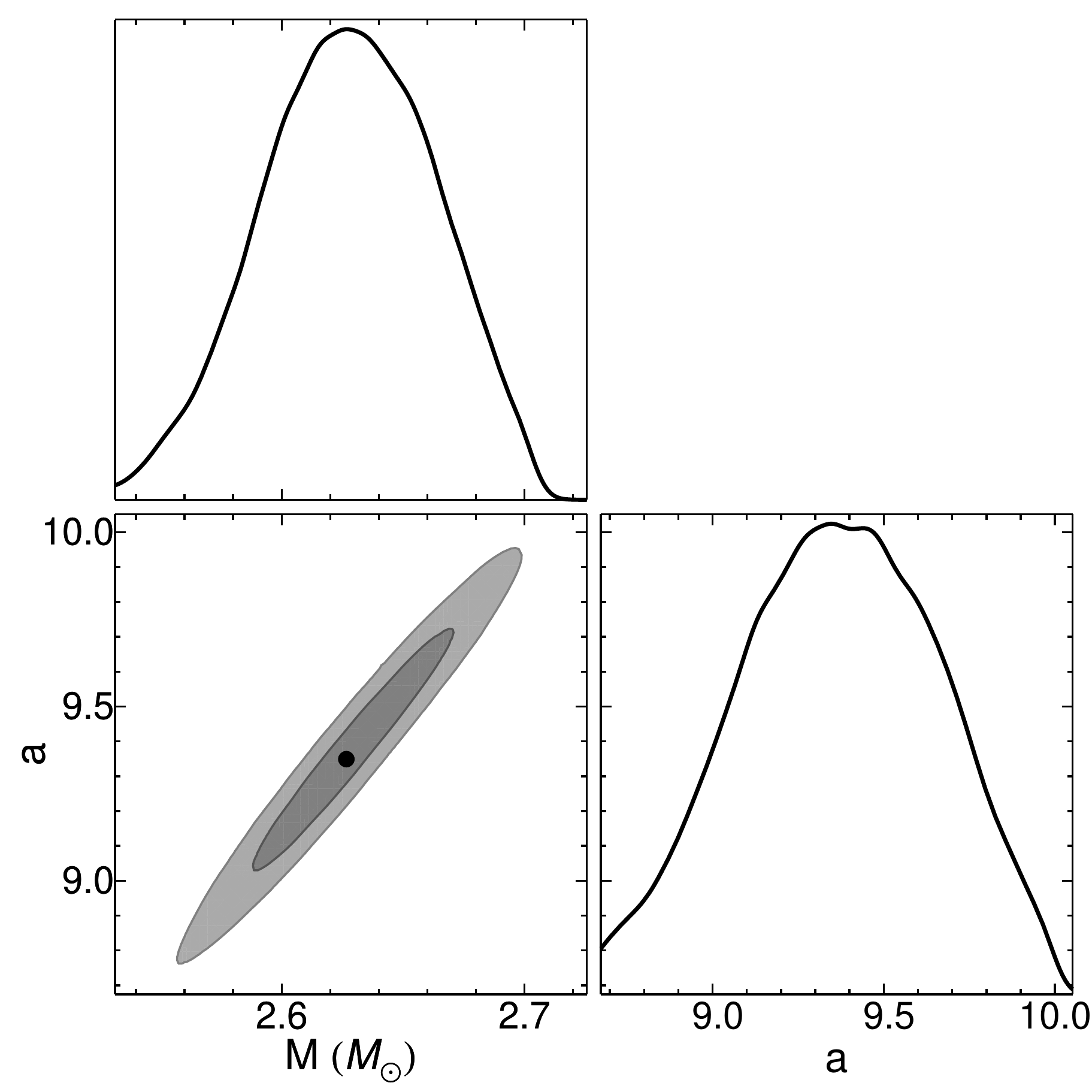}
\hfill
\includegraphics[width=0.24\hsize,clip]{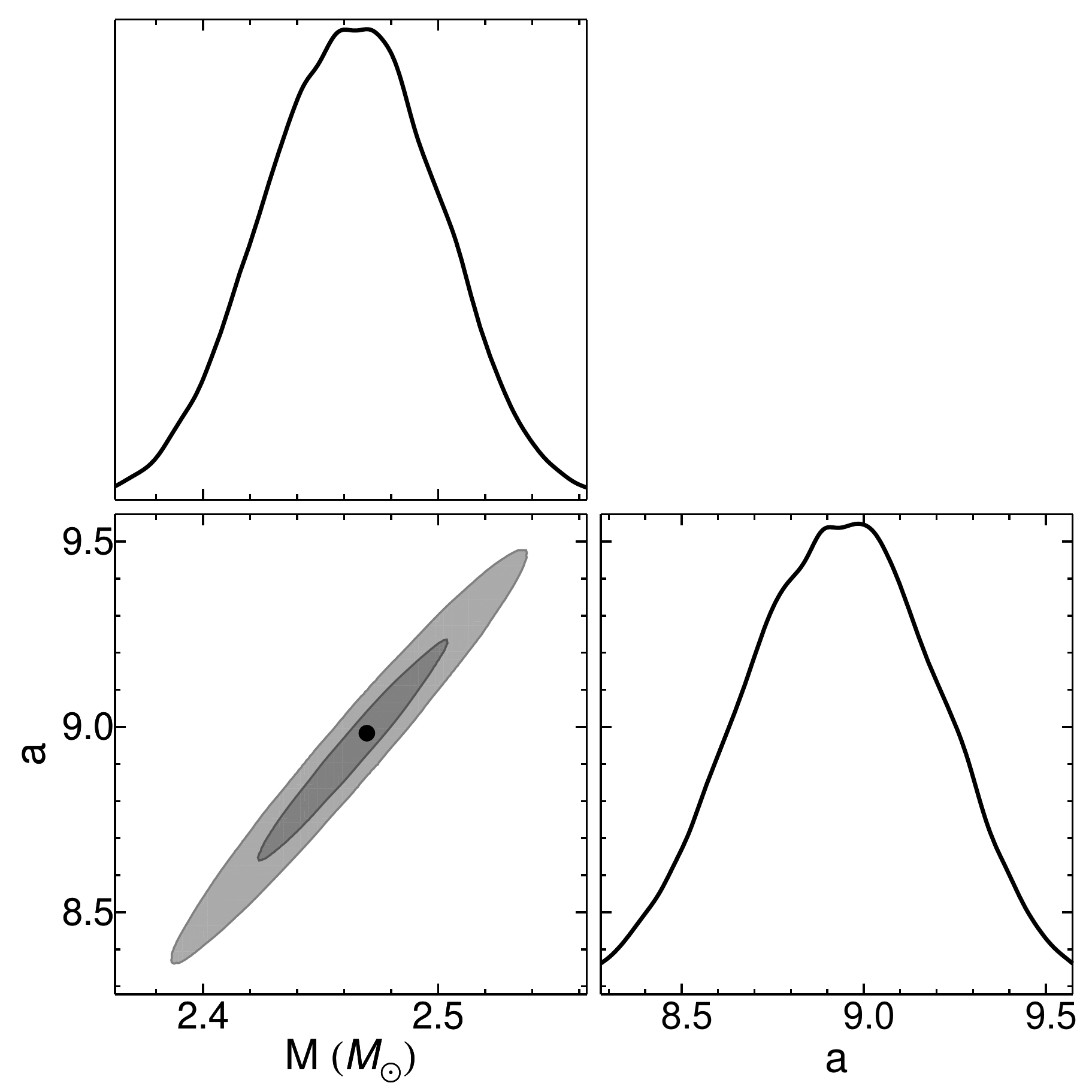}\hfill
\includegraphics[width=0.24\hsize,clip]{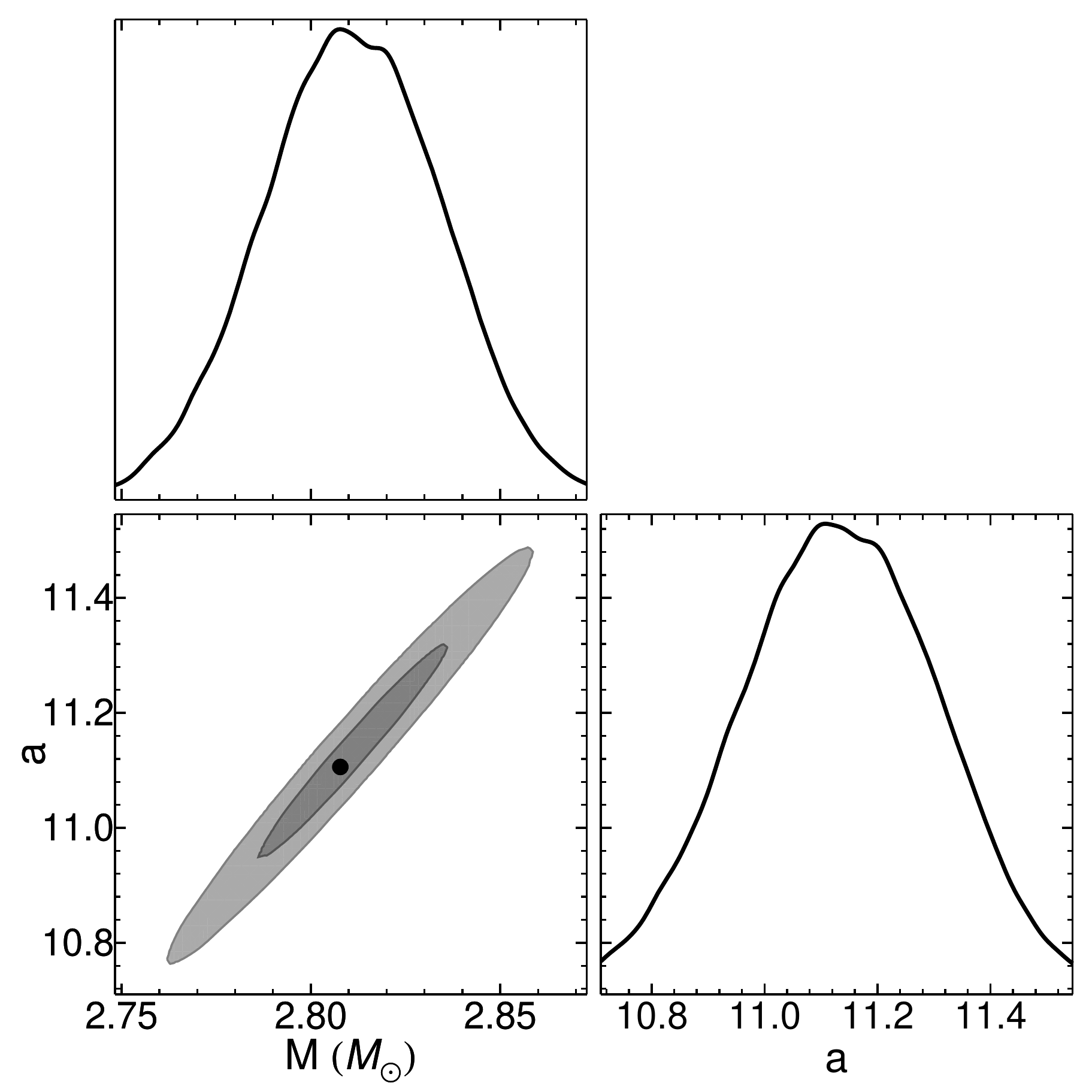}\hfill}
\caption{The Hayward metric contours plots of the best-fit parameters (black dots) and the associated 1--$\sigma$ (dark gray) and 2--$\sigma$ (light gray) confidence regions of the sources listed in Tab. \ref{tab:results}. Top panels, from left to right: Cir~X1, GX~5-1, GX~17+2, and GX~340+0. Bottom line, from left to right: Sco~X1, 4U1608-52, 4U1728-34, and 4U0614+091.}
\label{fig:contoursH}
\end{figure*}

\begin{figure*}[t]
{\hfill
\includegraphics[width=0.24\hsize,clip]{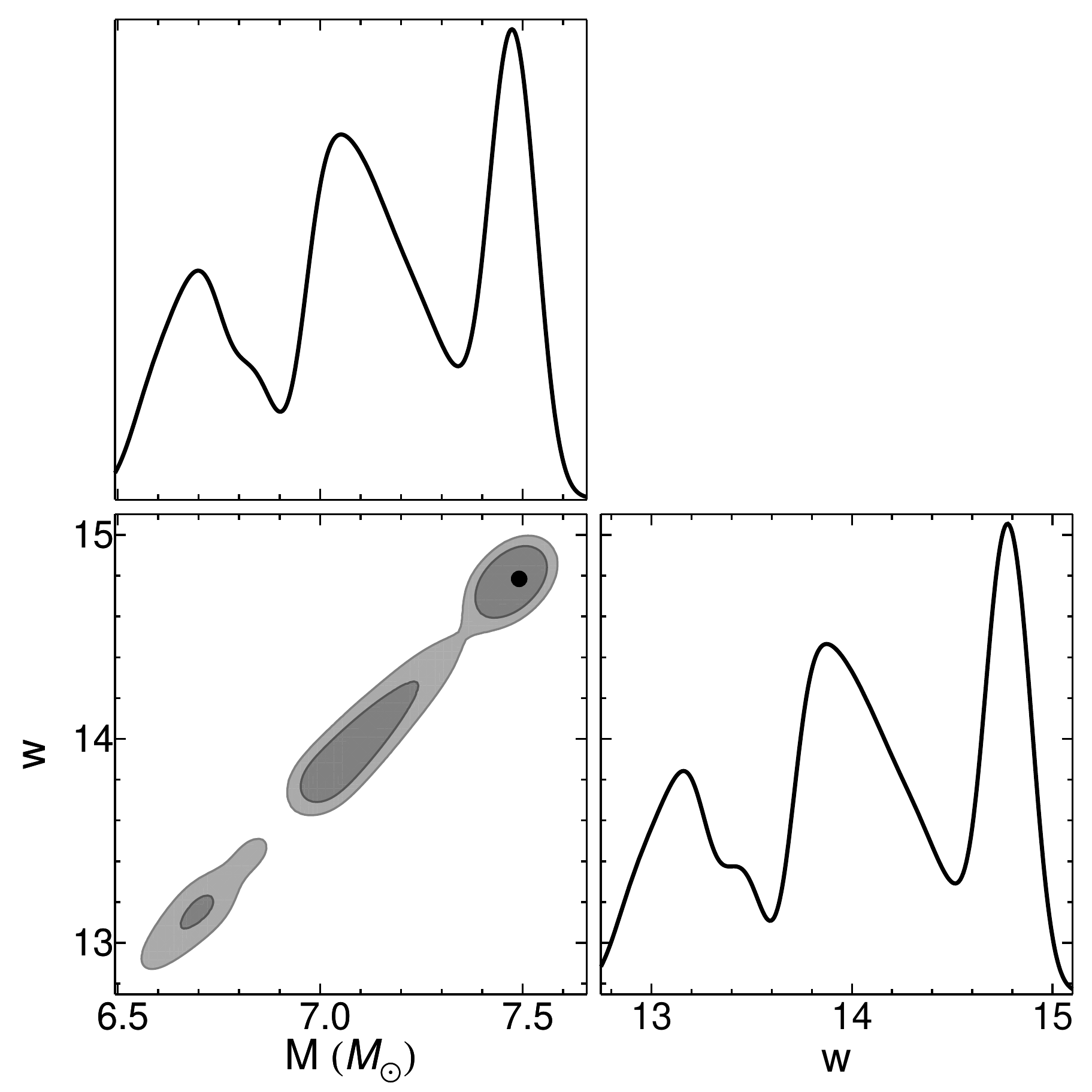}\hfill
\includegraphics[width=0.24\hsize,clip]{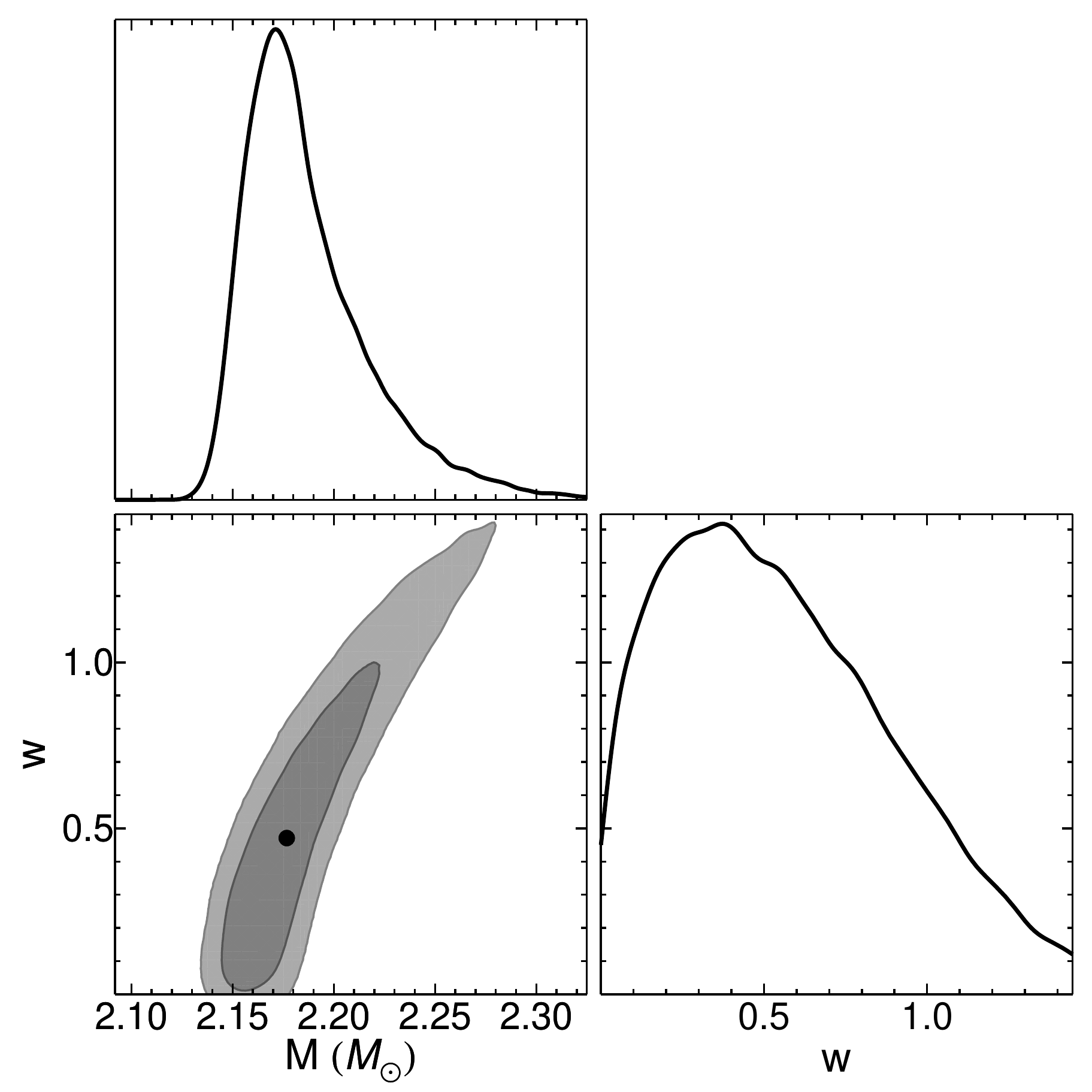}
\hfill
\includegraphics[width=0.24\hsize,clip]{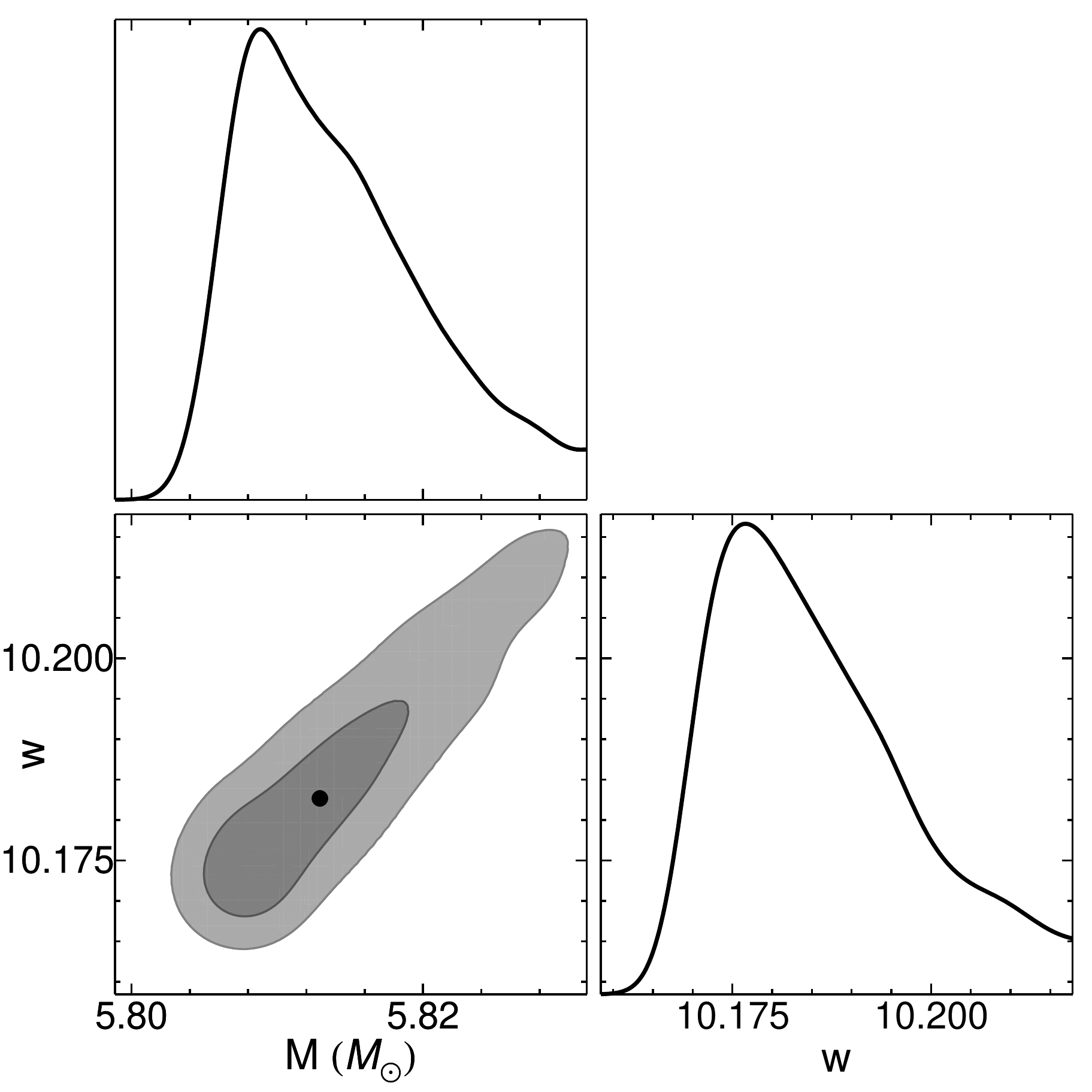}\hfill
\includegraphics[width=0.24\hsize,clip]{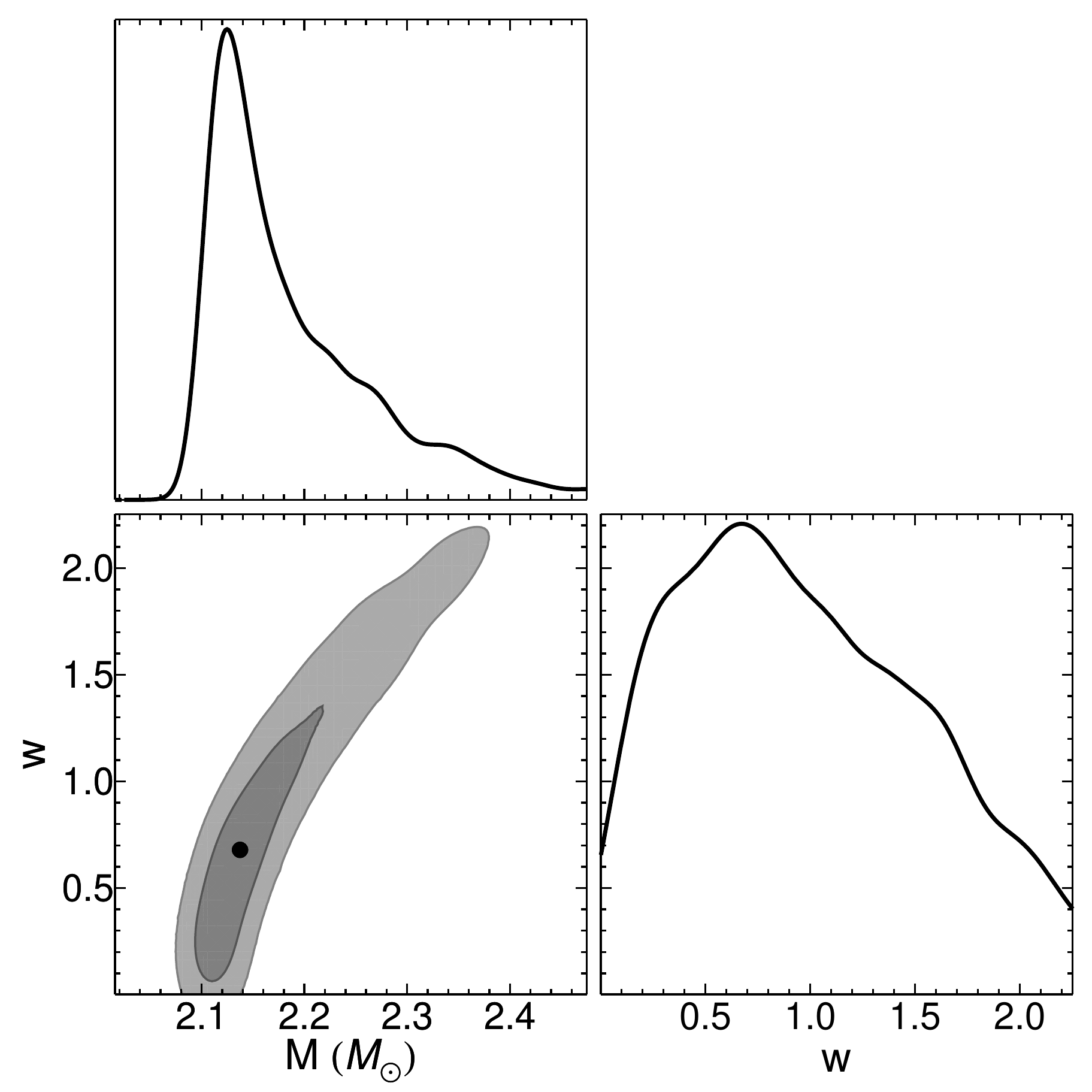}\hfill}

{\hfill
\includegraphics[width=0.24\hsize,clip]{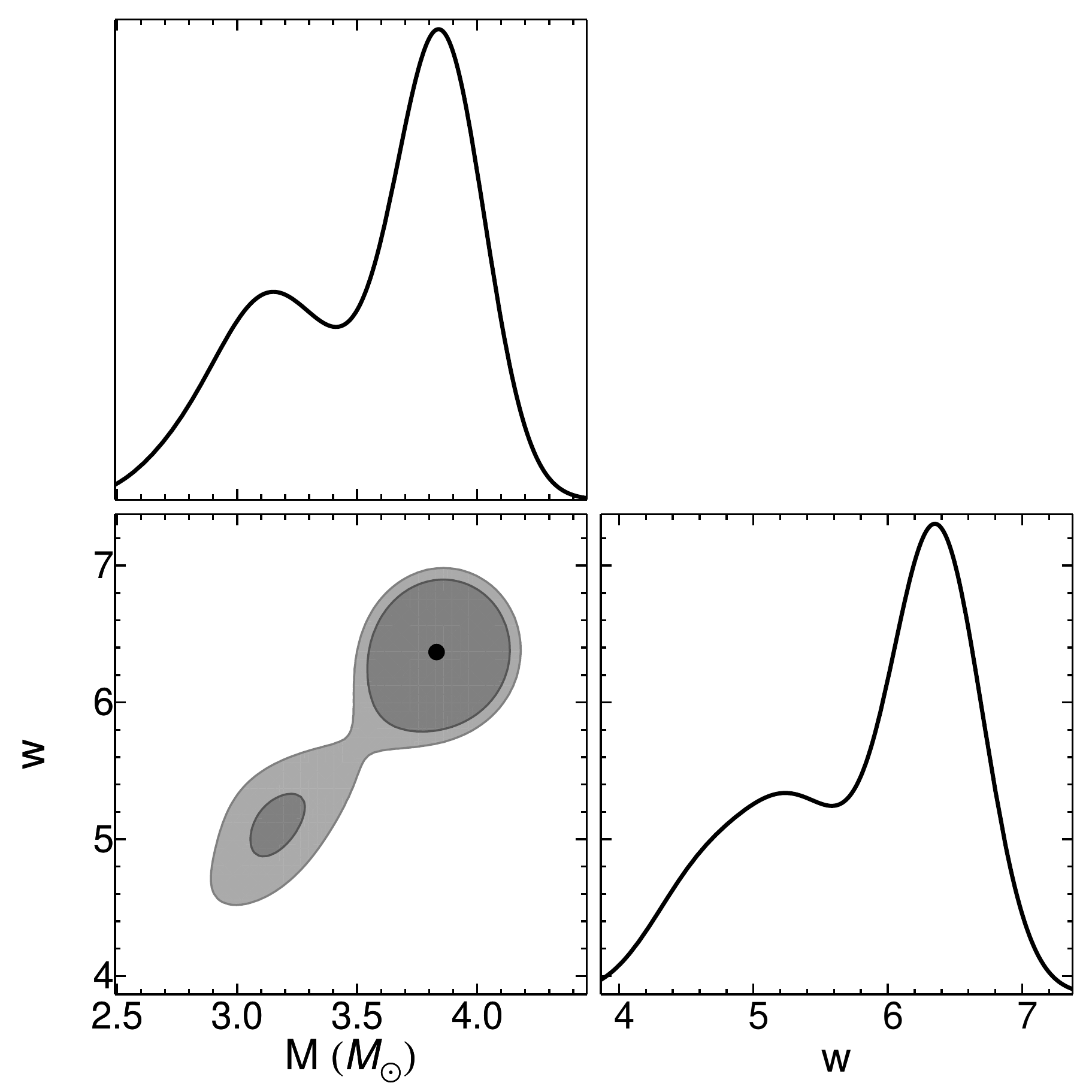}\hfill
\includegraphics[width=0.24\hsize,clip]{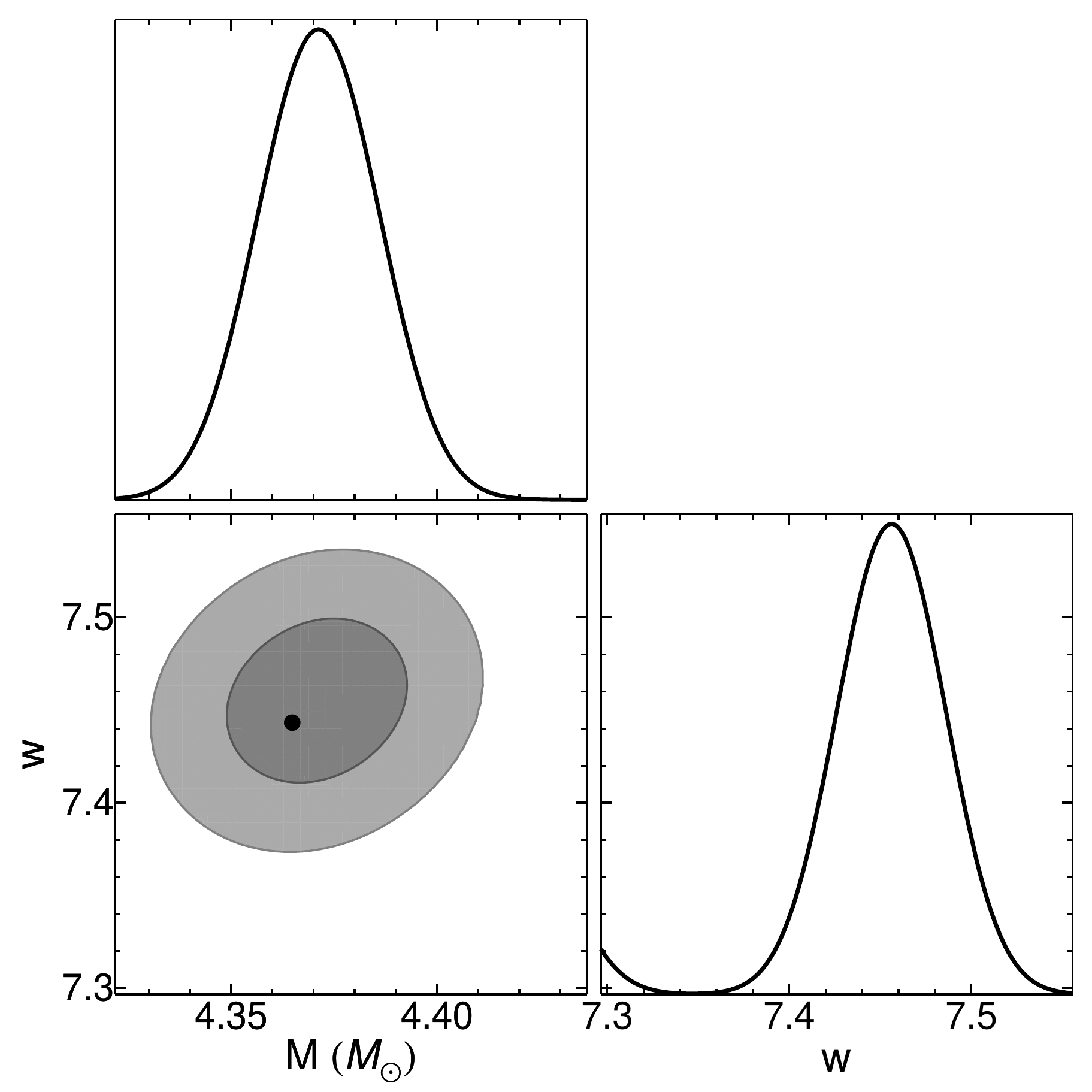}
\hfill
\includegraphics[width=0.24\hsize,clip]{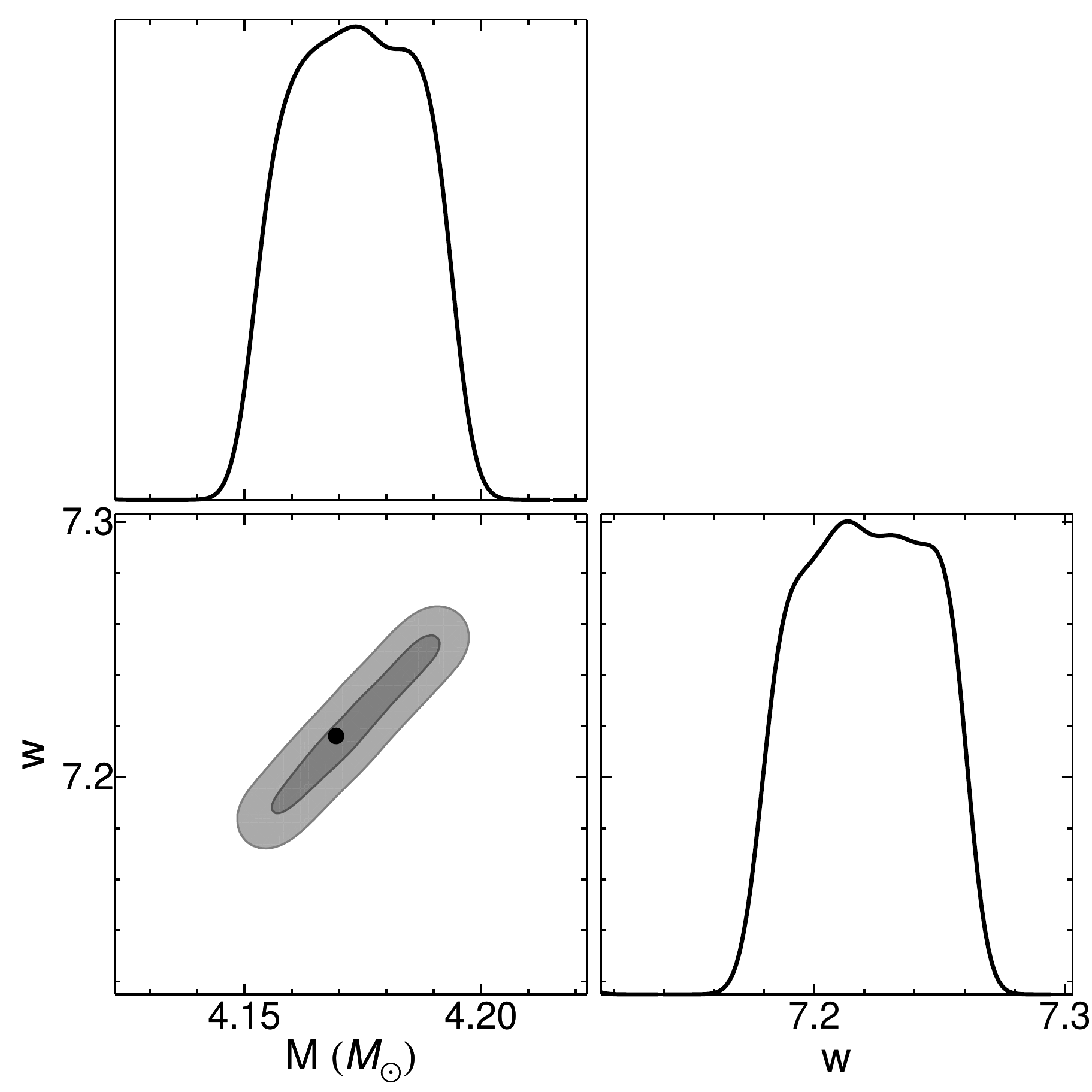}\hfill
\includegraphics[width=0.24\hsize,clip]{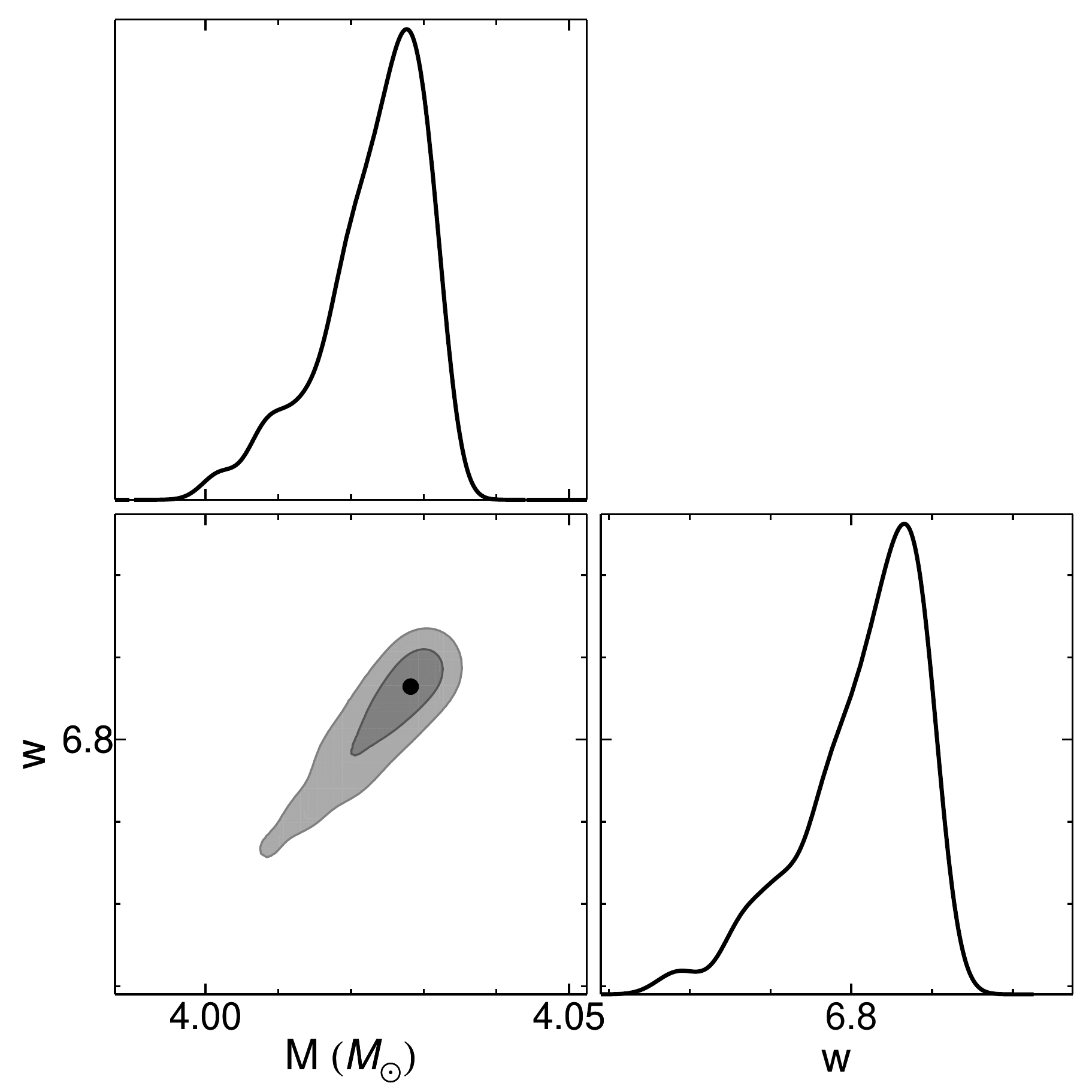}\hfill}
\caption{The same as in Fig.~\ref{fig:contoursH} but for Dymnikova metric. So that, the Dymnikova metric contours plots of the best-fit parameters (black dots) and the associated 1--$\sigma$ (dark gray) and 2--$\sigma$ (light gray) confidence regions of the sources listed in Tab. \ref{tab:results}. Top panels, from left to right: Cir~X1, GX~5-1, GX~17+2, and GX~340+0. Bottom line, from left to right: Sco~X1, 4U1608-52, 4U1728-34, and 4U0614+091.}
\label{fig:contoursD}
\end{figure*}

From the above procedure, using the data of each source we have obtained the results summarized in Tabs.~\ref{tab:results}--\ref{tab:isco}. In particular, Fig.~\ref{fig:freq} showcases the fitting of the $f_{\rm L}$--$f_{\rm U}$ QPO frequencies for the four models here-developed. Moreover, in Figs.~\ref{fig:contoursH}--\ref{fig:contoursD}, we display the contour plots of the best-fit model parameters listed in Tab.~\ref{tab:results}. The contours show the goodness of our findings up to 2--$\sigma$. As it appears evident, not all the fits smoothly converge as likely due to the lack of large data point number.  The corresponding $\sigma$ values are then affected by large error bars. This will reflect consequences on our theoretical conclusions, as discussed below.

\section{Theoretical discussion}

In this section, we argue some physical consequences based on the information summarized in Tabs.~\ref{tab:results}--\ref{tab:isco}. There, for each source, we can establish the best-fit model out of the four metrics considered throughout this work.

To compare each fit, based on different metric, we single out two main selection criteria, employing the well-consolidate \emph{Aikake Information Criterion} (AIC) and  \emph{Bayesian Information Criterion} (BIC) \cite{2007MNRAS.377L..74L}.

They read
\begin{subequations}
\begin{align}
{\rm AIC}&=-2\ln L+2p\,,\\
{\rm BIC}&=-2\ln L+p\ln N\,,
\end{align}
\end{subequations}
where the maximum value of the log-likelihood $\ln L$ is taken from our findings reported in Tab. \ref{tab:results}.

If we label with AIC$_0$ and BIC$_0$ the lowest values of the AIC and BIC tests, the model with these values is referred to as the fiducial (best-suited) model and the other models compare with the fiducial one via the difference $\Delta{\rm AIC/}\Delta{\rm BIC}={\rm AIC/BIC}-{\rm AIC}_0/{\rm BIC}_0$.
These differences provide evidence against the proposed model or, equivalently, in favor of the fiducial one, as follows
\begin{itemize}
    \item[-] $\Delta{\rm AIC}$ or $\Delta{\rm BIC}\in[0,\,3]$, weak evidence;
    \item[-] $\Delta{\rm AIC}$ or $\Delta{\rm BIC}\in (3,\,6]$, mild evidence;
    \item[-] $\Delta{\rm AIC}$ or $\Delta{\rm BIC}>6$, strong evidence.
\end{itemize}

For each QPO source, we discuss the results from statistical, observational and physical points of view. In what follows, we want to stress that, though depending on the equation-of-state, theoretically one can have larger masses up 6.1 M$_\odot$ \cite{2002BASI...30..523S}, albeit from observations we now get up to 2.14 M$_\odot$ \cite{2020NatAs...4...72C}. We will \emph{disentangle} this concept in our theoretical interpretations below.

\begin{itemize}
\item[-] {\bf Cir~X-1} \cite{2006ApJ...653.1435B}.
From Tab.~\ref{tab:results}, we notice that Bardeen and Dymnikova metrics are equally strongly preferred over the other ones and, from Tab.~\ref{tab:isco}, these RBH solutions provide suitable physical values for the ISCO.
However, in both cases we obtain masses $M\gtrsim 3$~M$_\odot$, which seems incompatible with the NS interpretation supported by Ref.~\cite{2010ApJ...714..748T}.
However, if we consider the absolute upper limit of $M_{\rm up}=6.1$~M$_\odot$ \cite{2002BASI...30..523S}, Bardeen RBH provides a mass below $M_{\rm up}$, whereas Dymnikova RBH does not.
Therefore, we conclude that Bardeen RBH is the favored solution from the statistical, theoretical and experimental perspectives.
\item[-] {\bf GX~5-1} \cite{1998ApJ...504L..35W,2002MNRAS.333..665J}, {\bf GX~17+2} \cite{2002ApJ...568..878H}, {\bf GX~340+0} \cite{2000ApJ...537..374J}.
Looking at Tabs.~\ref{tab:results}--\ref{tab:isco} for the source GX~5-1, we notice that all models do provide good fits to the data and physical values for the ISCO. From a statistical viewpoint, the Schwarzschild metric with its parameters represents the fiducial model that provides a well constrained mass, compatible with current NS mass observations. Hence, we conclude that it represents the best-fit model.
Similar conclusions, but with stronger evidences against the more complicated models, can be reached also for the source GX~340+0, where all the RBHs are statistically disfavored.
On the contrary, in the case of GX~17+2, Hayward and Bardeen metrics are equally good fits to data (see Tab.~\ref{tab:results}). The Hayward metric provides a mass which is barely consistent with the NS interpretation, if one considers extremely rotating NSs with stiff equations-of-state \cite{2015PhRvD..92b3007C}, while the mass inferred from the Bardeen metric is not consistent at all with any NS observations, albeit theoretically not fully-excluded \cite{2002BASI...30..523S}.
On the ground of these considerations, the Hayward and Bardeen metrics provide the best-fitting and physically-allowed solutions that correctly describe GX~17+2.
\item[-] {\bf Sco~X1} \cite{2000MNRAS.318..938M}.
From Tab.~\ref{tab:results}, the best fit is given by the Bardeen metric, mildly preferred over the Hayward one. Even in this case, the main caveat is related to the mass: Bardeen metric provides a mass which is barely consistent with that of a very extreme NS \cite{2015PhRvD..92b3007C}, while the mass $\approx2.7$~M$_\odot$ obtained from the Hayward metric is in the acceptable observational range \cite{2014NuPhA.921...33B}. Moreover, from Tab.~\ref{tab:isco} all the RBH metrics do provide physical ISCO values.
Therefore, the Bardeen and Hayward metrics do give the only physical fit for this source.
It is worth mentioning that the masses inferred from the Bardeen and Hayward metrics (see Tab.~\ref{tab:results}) are inconsistent with the range $1.40$--$1.52$~M$_\odot$, obtained by modelling optical light curves of Sco~X1 \cite{2021MNRAS.508.1389C}. This fact suggests that all the considered metrics do not accurately describe the QPO data of Sco~X1, therefore, further analyses will be performed in future works.
\begin{figure*}[t]
{\hfill
\includegraphics[width=0.24\hsize,clip]{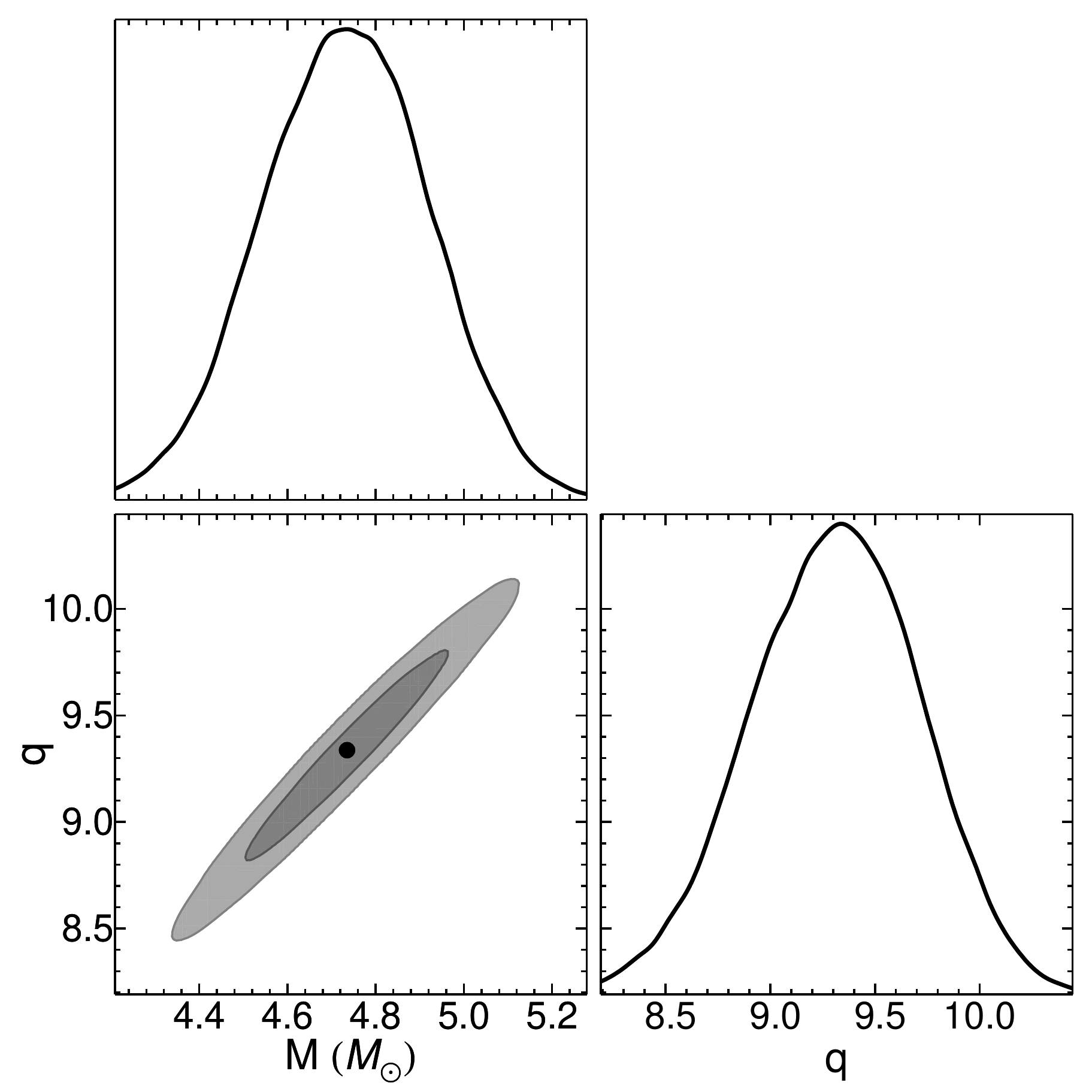}\hfill
\includegraphics[width=0.24\hsize,clip]{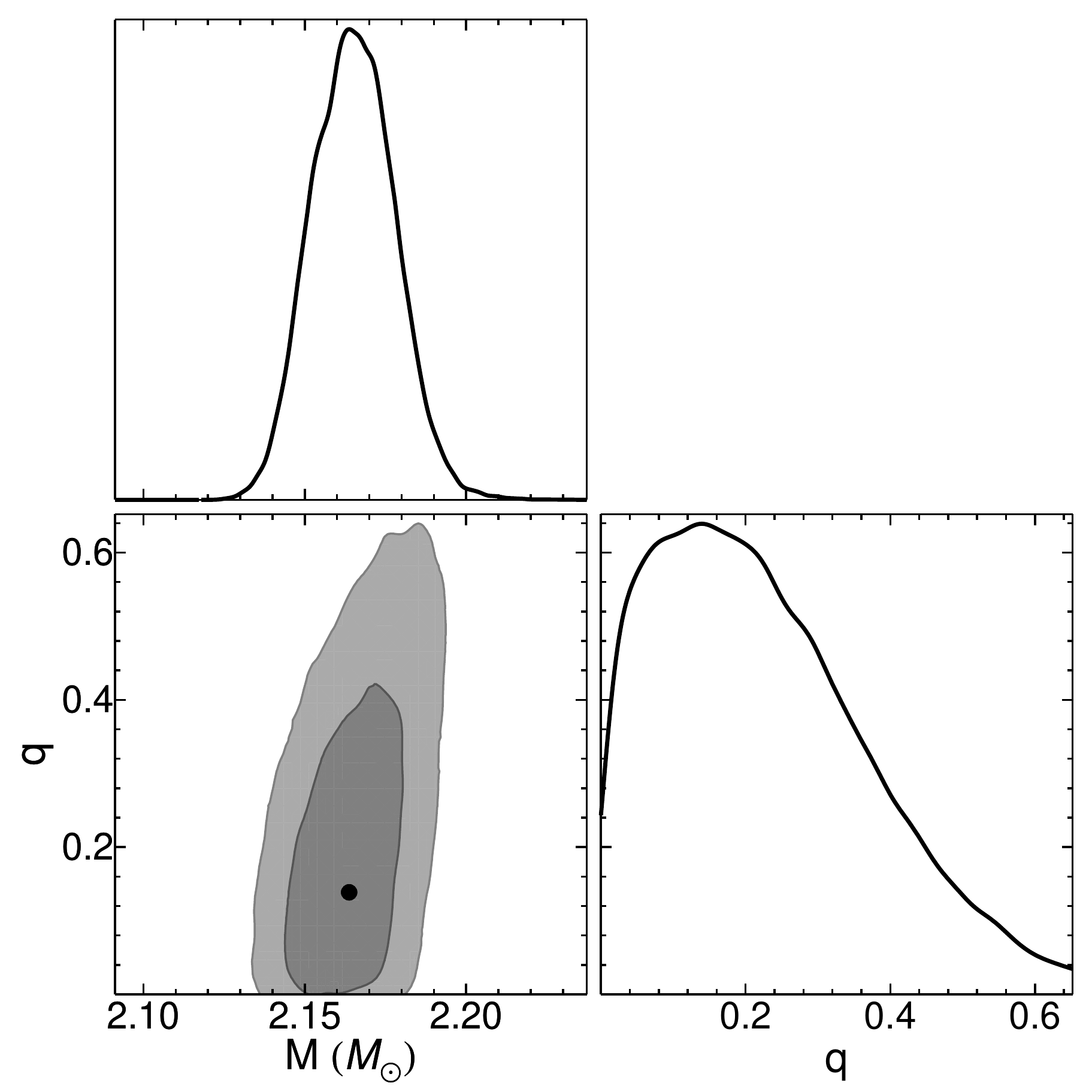}
\hfill
\includegraphics[width=0.24\hsize,clip]{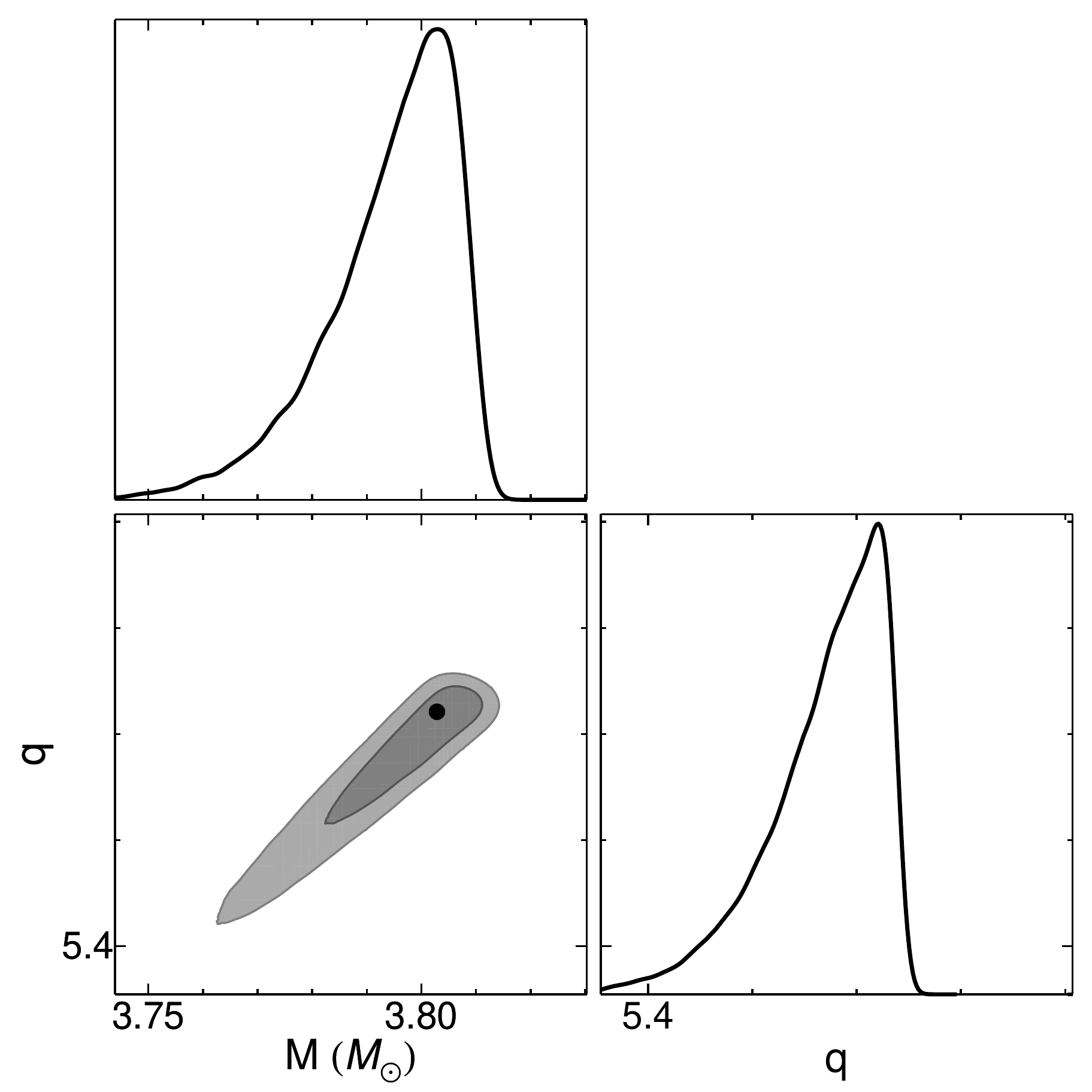}\hfill
\includegraphics[width=0.24\hsize,clip]{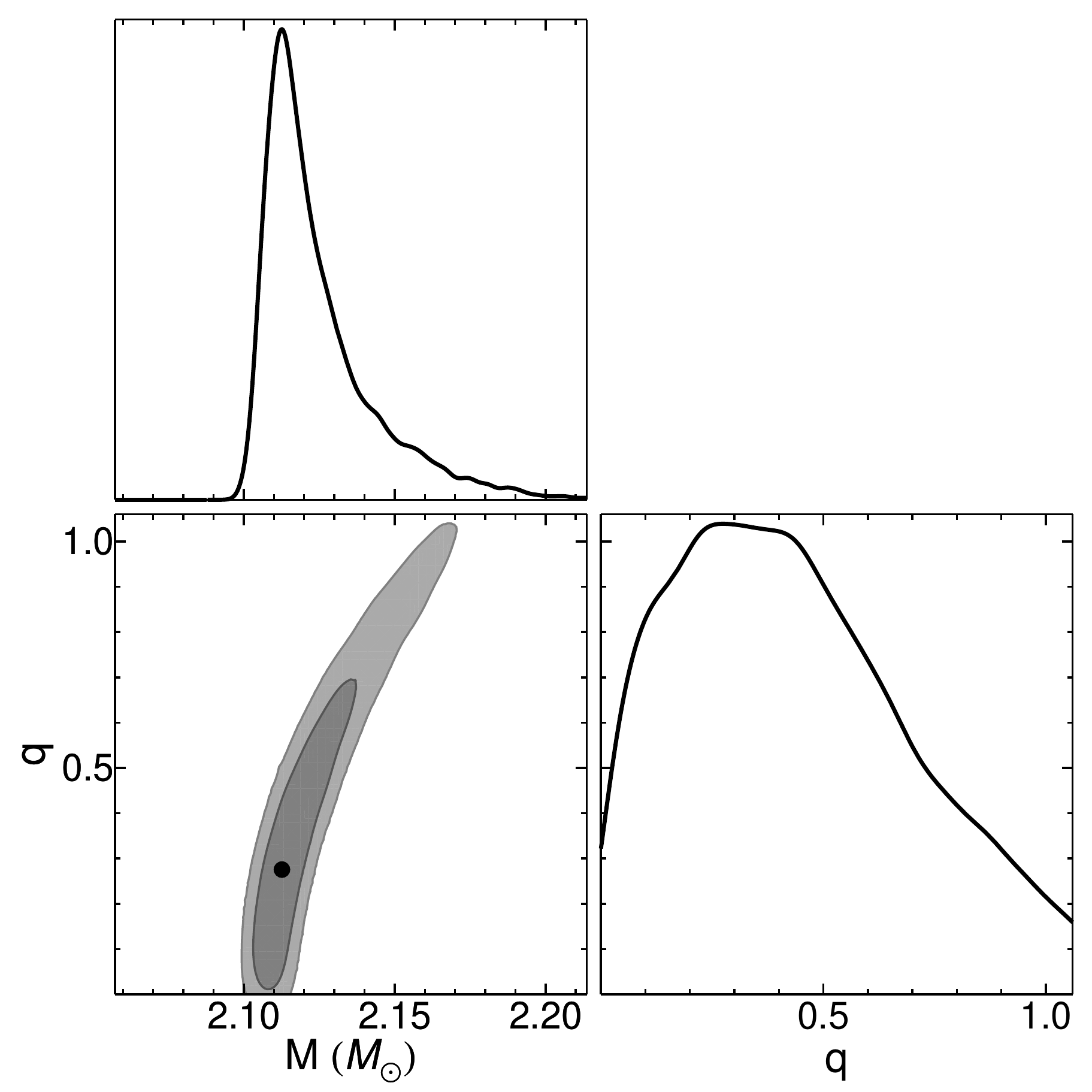}\hfill}

{\hfill
\includegraphics[width=0.24\hsize,clip]{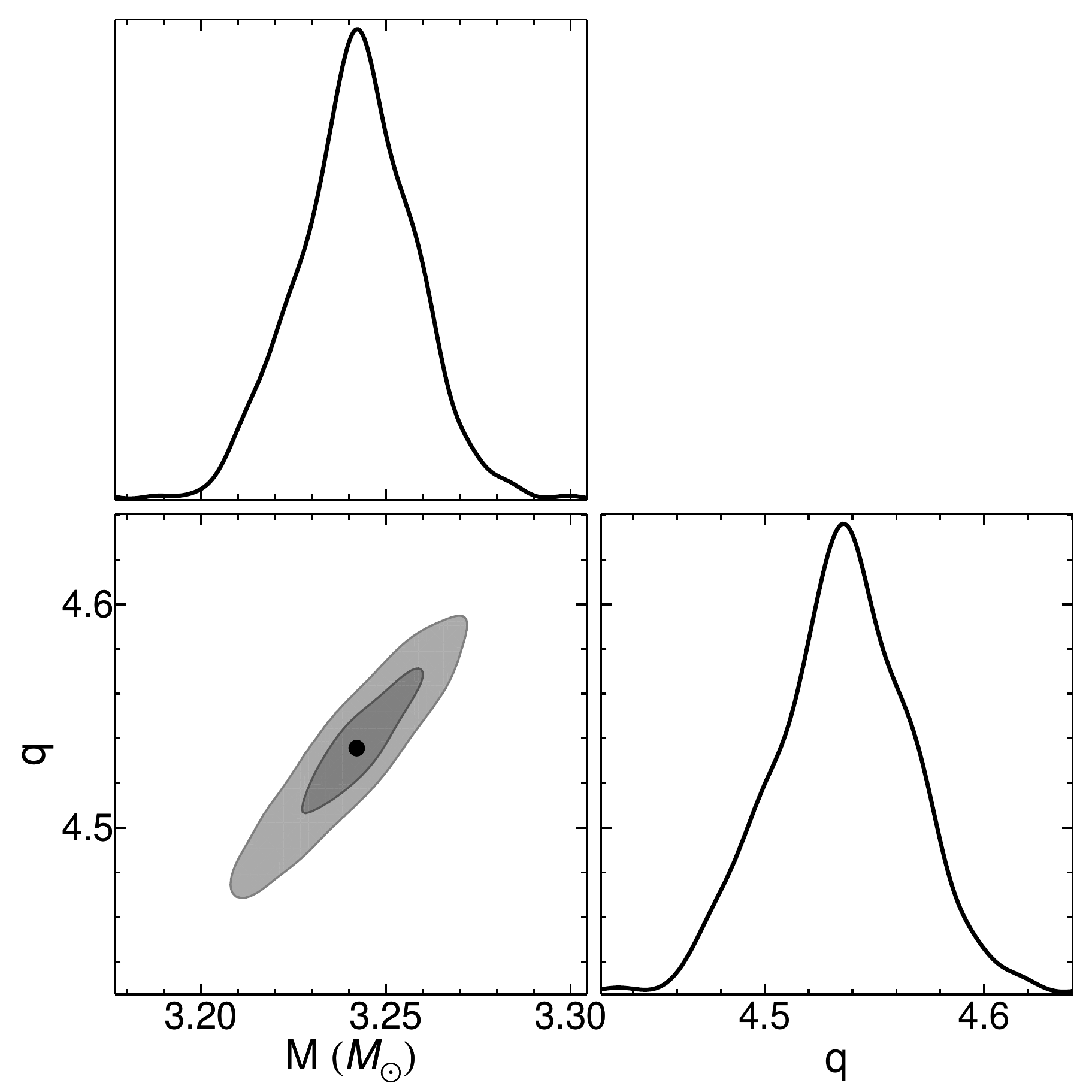}\hfill
\includegraphics[width=0.24\hsize,clip]{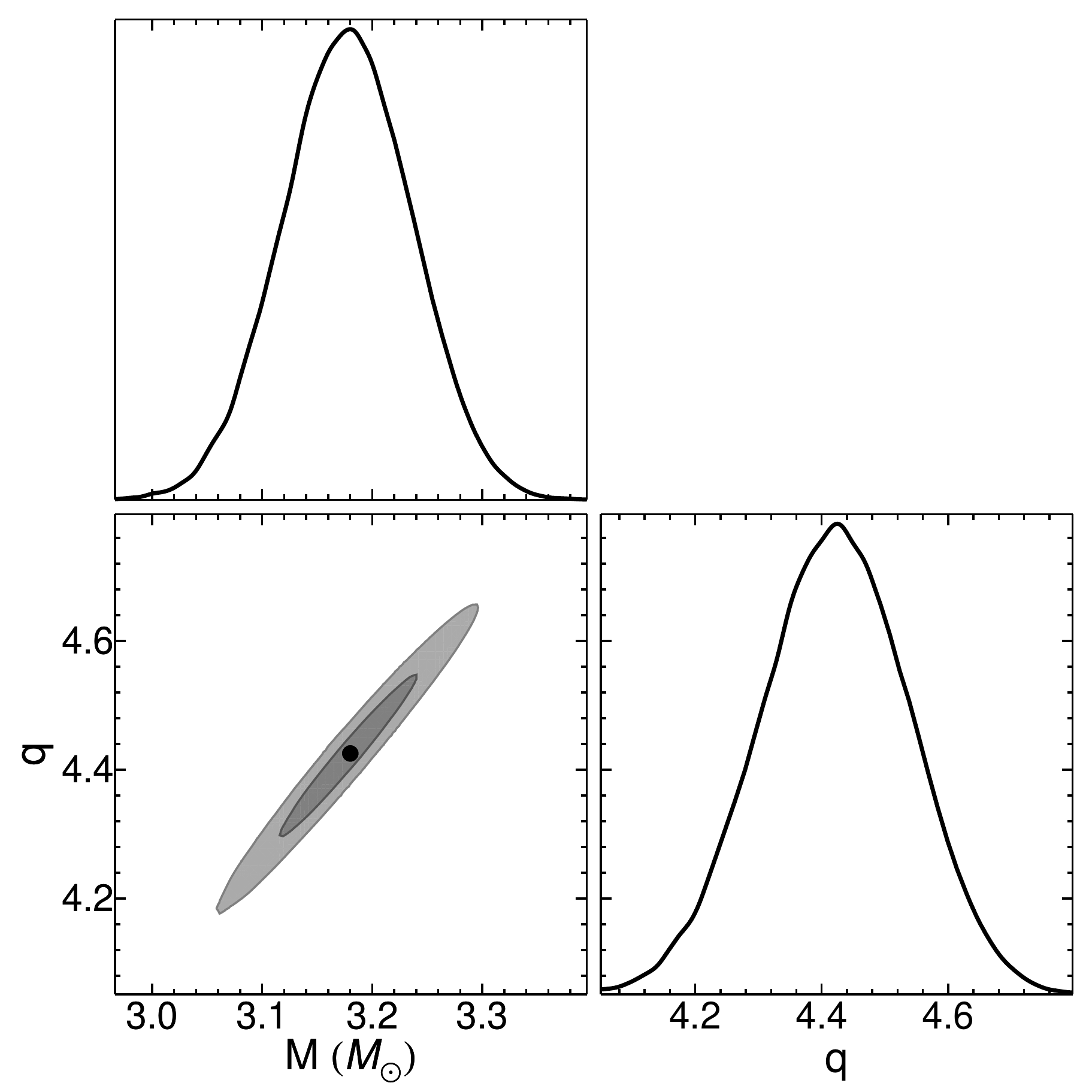}
\hfill
\includegraphics[width=0.24\hsize,clip]{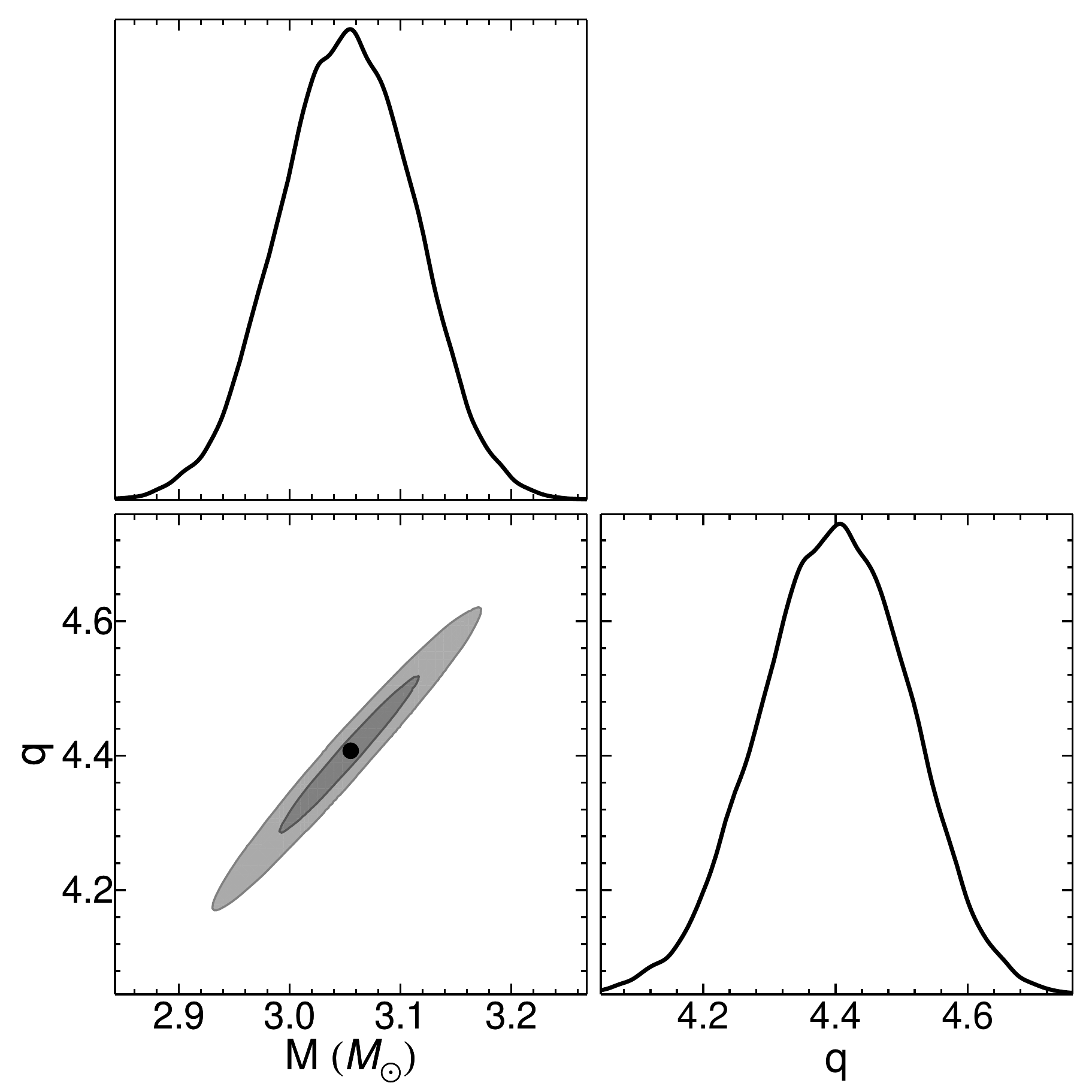}\hfill
\includegraphics[width=0.24\hsize,clip]{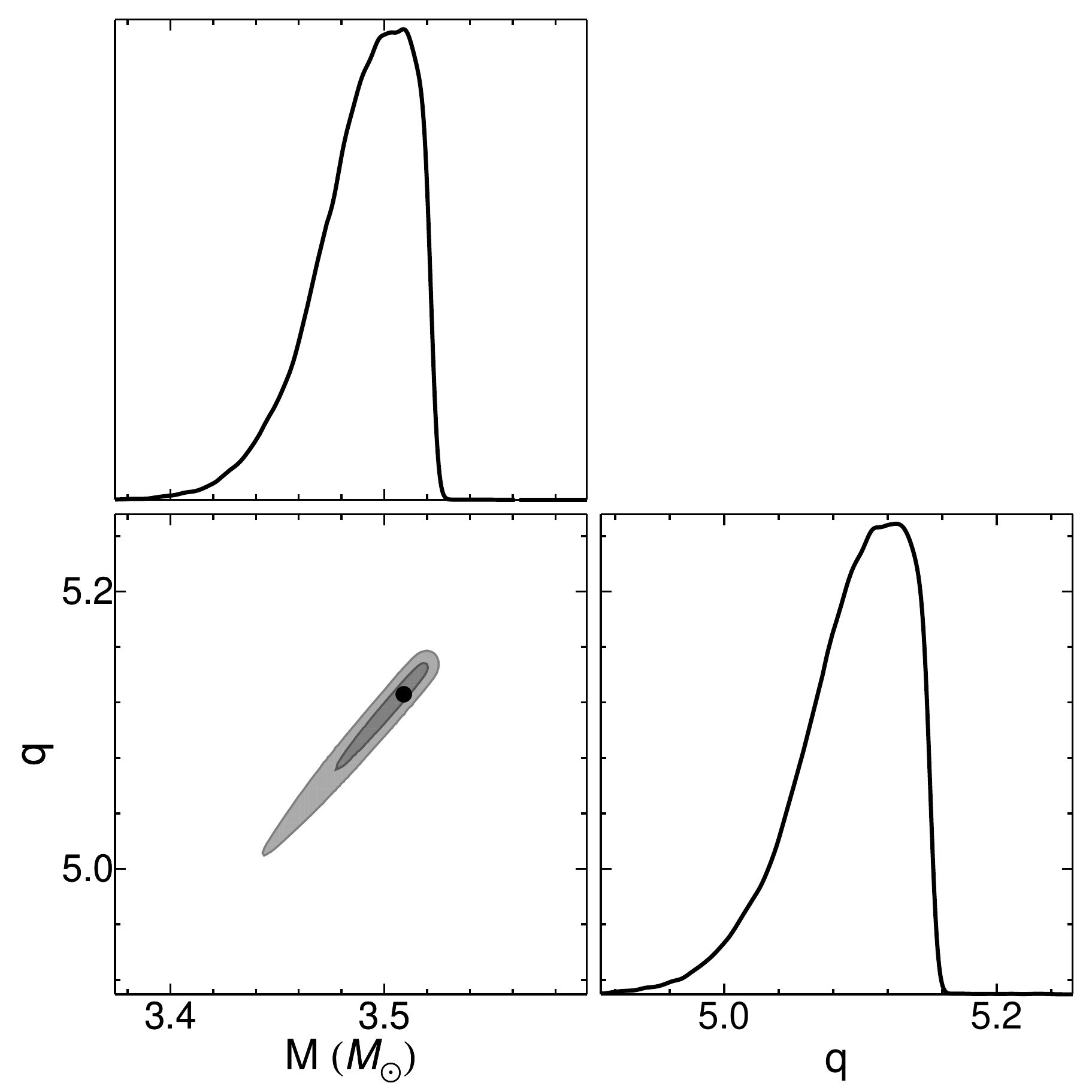}\hfill}
\caption{The same as in Fig.~\ref{fig:contoursH} but for the Bardeen metric.}
\label{fig:contoursB}
\end{figure*}
\item[-] {\bf 4U1608-52} \cite{1998ApJ...505L..23M}, {\bf 4U1728-34}  \cite{1999ApJ...517L..51M}, {\bf 4U0614+091} \cite{1997ApJ...486L..47F}.
The considerations reached for 4U1608-52 are analogous to the previous case of Sco~X1, with the best fit given by the Bardeen metric slightly preferred over the Hayward one.
Again, the Hayward metrics provide a mass which is not as extreme as for the Bardeen metric, all of them theoretically not fully-excluded \cite{2002BASI...30..523S}.
For 4U1728-34, the Bardeen and Dymnikova metrics perform equally well, but Bardeen metric does provide an extreme NS mass, still consistent with the upper limit from  Ref.~\cite{2021MNRAS.508.1389C}, whereas Dymnikova does not, albeit still below $M_{\rm up}$.
For 4U0614+091, the best fit model is obtained unambiguously from the Hayward metric because it is the fiducial model and from a physical point of view provides both a NS-like mass and a physical values of the ISCO.
\end{itemize}

We thus can finally conclude that, guided by the NS physical interpretation of mass constraints and by the values obtained for the ISCOs, the Schwarzschild metric \emph{is not the only physically-allowed solution} and turns out to be disfavored for most of the above listed QPO sources. For all the sources, the Bardeen and Hayward solutions are successful to model the QPO frequencies also from a statistical viewpoint, see Figure~\ref{fig:freq}.

For the sake of completeness, it appears that the contour plots obtained from RBHs are sometimes badly shaped.
This is particularly evident for the Dymnikova, metric, which appears unsuitable to model QPOs. In view of the above, we conclude that singular spacetimes seem less predictive to frame out the exteriors of a NS, and so finding out new regular spacetimes, \emph{ad hoc} constructed to mode such sources, could represent a future key perspective. In other words, we may speculate that new RBH solutions can better adapt to these kind of problems as we will investigate in incoming efforts.

Last but not least, following Ref.~\cite{2022arXiv221210186B}, we see that, among all singular metrics, the de Sitter and/or anti-de Sitter solutions appear to work better in featuring the fits for our sources compared with the Schwarzschild metric. As it is well-known, the de Sitter and anti-de Sitter solutions prompt a cosmological constant term, $\sim \Lambda r^2$, filling the whole spacetime (not only at the center, as in the Hayward picture). The latter term, then, is clearly regular, supporting more our conclusions that seem to favor RBHs.

\section{Conclusions and perspectives}\label{sezione5}

We here focused on eight NS sources, fitting the corresponding observed frequency data with four QPO models, within the framework of the relativistic precession model. In this respect, we considered three RBH solutions, involving spherical symmetry: Bardeen, Hayward and Dymnikova spacetimes. In addition to these three, for comaprison, we cosidered also the standard Schwarzschild solution. In so doing, we analyzed the aforementioned RBH solutions by means of MCMC fitting procedure based on the Metropolis-Hastings algorithm. Then, our findings have been compared and contrasted by adopting AIC and BIC statistical criteria. Consequently, for each QPO data set, we computed best fit values, in particular the masses, and inferred the ISCO values for each solution.
Our results certify that RBHs can describe NS exteriors and in most of the cases appear to be better suited than the standard spherical symmetry induced by the Schwarzschild spacetime. To this end, we conclude that among all the RBH involved solutions, the best options remain Bardeen and Hayward spacetimes, whereas the Dymnikova metric is ruled out.

This consideration could be justified noticing that the RBH solutions required topological charges and/or effective cosmological constant behavior at the center that seem to be relevant to characterize the external of a compact object. Our results, in view of recent findings, suggested that a possible de Sitter and anti-de Sitter core-solutions better fit our sources \cite{2022arXiv221210186B} and the possibility of extending with non-singular metrics remains fully-valid. As direct drawback, we can stress the need of more and better QPO data sets to improve the overall quality of the fits. However, we are convinced that our findings will be compatible with actual results, confirming the Bardeen and Hayward spacetimes.

As future perspectives, we will focus on additional regular metrics. Moreover, we will reconstruct a possible regular approach based on data, adopting some sort of \emph{back-scattering procedure} of reconstruction. In addition, we will test the RBHs with frameworks different from the relativistic precession model.

Future efforts will focus on studying more RBH solutions that exhibit physical properties different from topological charges. The same procedure can be carried forward involving alternative theories of gravity where RBHs can be investigated, checking whether changing the background is compatible with NS sources.  Finally, we will investigate the effects of reconstructed non-singular metrics, i.e., regular solutions inferred directly from data, instead of postulating a given one from the very beginning.

\section*{Acknowledgements}
OL is grateful to the Department of Physics of the
Al-Farabi University for hospitality during the period in which this manuscript has been written. OL acknowledges Roberto Giamb\`o for fruitful discussions on the subject of this paper. KB acknowledges Mariano Mendez for providing QPO data. This research has been partially funded by the Science Committee of the Ministry of Science and Higher Education of the Republic of Kazakhstan (Grant No. AP19680128).

\end{document}